\documentclass[a4paper,fleqn,usenatbib]{mnras}
\usepackage{epsfig}
\usepackage{graphicx,wrapfig,lipsum}	
\usepackage{amsmath}     
\usepackage{hyperref}	    
\usepackage{amssymb}   	
\usepackage{multicol}    
\usepackage{multirow}
\usepackage{bm}		    
\usepackage{pdflscape}

\usepackage[T1]{fontenc}
\usepackage{ae,aecompl}
\usepackage{txfonts}
\usepackage{array}
\usepackage{caption}
\usepackage{threeparttable}
\title[Physical Properties of M$-$dwarfs]{Understanding Physical Properties of Young M-dwarfs: NIR spectroscopic studies.}
\author[Khata et al. 2018]{
Dhrimadri Khata,$^{1}$\thanks{E-mail: dhrimadrikht@gmail.com (DK)}
Soumen Mondal,$^{1}$
Ramkrishna Das,$^{1}$
Supriyo Ghosh$^{1,2}$
and Samrat Ghosh$^{1}$
\\
$^{1}$Satyendra Nath Bose National Centre for Basic Sciences,
Block-JD, Sector-III, Salt Lake, Kolkata-700 106\\
$^{2}$ Tata Institute of Fundamental Research, Mumbai 400005, India
}

\date{Accepted XXX. Received YYY; in original form ZZZ}

\pubyear{2018}

\begin{document}
\label{firstpage}
\pagerange{\pageref{firstpage}--\pageref{lastpage}}
\maketitle

\begin{abstract}
We present here medium resolution ($\lambda /\Delta\lambda\sim$1200) H- and K-band spectra of M type dwarf stars covering the wavelength range 1.50 - 1.80 $\mu$m and 1.95 - 2.45 $\mu$m. The sample includes 53 dwarf stars (M0V-M7V) from new observations using the TIFR Near-Infrared Spectrometer and Imager (TIRSPEC) instrument on the 2-m Himalayan Chandra Telescope (HCT). Using interferometrically-measured effective temperature ($T_{eff}$), radius and luminosity of nearby bright calibrator stars, we have created new empirical relationships among those fundamental parameters and spectral indices. The equivalent widths of H-band spectral features like Mg (1.57 $\mu$m), Al (1.67 $\mu$m) and Mg (1.71 $\mu$m), and the $H_{2}O$-H index are found to be good indicators of $T_{eff}$, radius and luminosity and we establish the linear functions using these features relating to those stellar parameters. The root mean squared error (RMSE) of our best fits are 102K, 0.027$R_{\sun}$ and 0.12dex respectively. Using spectral type standards along with known parallaxes, we calibrate both H and K-band $H_{2}O$ indices as a tracer of spectral type and absolute $K_{s}$ magnitude. Metallicities of M-dwarf samples are estimated using the K-band calibration relationships. The mass of M-dwarfs could be determined using the luminosity (L/$L_{\sun}$) and we establish a new empirical relation for this. We also compare and contrast our results with other similar work from the literature. 
\end{abstract}

\begin{keywords}
stars:fundamental parameters -- stars:M-dwarfs -- techniques:spectroscopic.
\end{keywords}



\section{Introduction :}

$~~~~~$ M-dwarf stars within our galaxy are more than 70$\%$ of all stars (Henry et al. \citet{Henry2006}), dominating the stellar populations by number (Bastian et al.
\citet{Bastian2010}), having a very low mass range (0.075$M_\odot$-0.50$M_\odot$, Delfosse et al. \citet{Delfosse2000}) and an effective temperature ($T_{eff}$) less than 4000K. Observational evidence confirms that the chances of the occurrence of planetary systems, especially earth-like planets orbiting in "habitable zones", increases with decreasing stellar mass and radius (Howard et al. \citet{Howard2012}). Due to their proximity, small size, and low mass, M-dwarfs are becoming attractive targets for potentially habitable exoplanet searches via almost all current search methods. The main advantages of choosing M dwarfs are the deeper transit depths which helps to detect planet's observable signals easily for a given planet radius and the closeness of the habitable zones to the host stars making it geometrically favorable for observing the transit and increasing the frequency of transits. M dwarfs also represent a complete archaeological record of the chemical evolution and star formation history of the milky way galaxy (Bochanski et al. \citet{Bochanski2007}). \\
$~~~~~~~$The characterization of low-mass stars is very crucial, and NASA's Kepler mission suggests that M dwarfs are swarming with rocky planets (Fressin et al. \citet{Fressin2013}), by looking for planet transits and through the identification of thousands of planet candidates (Borucki et al. \citet{Borucki2010}, \citet{Borucki2011a}, \citet{Borucki2011b}; Batalha et al. \citet{Batalha2013}; Burke et al. \citet{Burke2014}). Mid to late M dwarfs are considered to be appropriate targets for transiting planet searches (Nutzman and Charbonneau \citet{Nutzman2008}), and recent findings from Kepler showed that there is less chance of giant planets being found around K and early M dwarfs than around F and G stars (Borucki et al. \citet{Borucki2011a}; Fressin et al. \citet{Fressin2013}). While Stellar characterization for these kinds of cool dwarfs is a notoriously difficult problem, the small radius of the M dwarfs enables the orbiting planet's atmosphere to be studied with transmission or occultation techniques and precise spectroscopic studies of nearby bright M dwarfs are possible using near-infrared (NIR) observations (Bean et al. \citet{Bean2011}; Crossfield et al. \citet{Crossfield2011}; Berta et al. \citet{Berta2012}). \\
$~~~~~~$It is very important to estimate the fundamental stellar parameters with precise accuracy and characterize these kinds of cooler dwarf stars both observationally and with theoretical perspectives. Uncertain sources of opacity and the mixture of complex molecules and grains in the M dwarfs atmosphere hampers the interpretation of observables, and the lack of accurate modeling of the deep convective zones in M dwarf interiors (Mullan and MacDonald \citet{Mullan2001}; Browning \citet{Browning2008}) and mismatch of oscillator strengths are the constraints of theoretical model prediction. In this scenario, empirical calibrations become strikingly useful and provides an inroad for determining the physical properties of M dwarfs. For example; Delfosse et al. (\citet{Delfosse2000}) estimates the mass using the Mass-$M_{k}$ relation for stars with known parallaxes, others (e.g., Bayless and Orosz \citet{Bayless2006}; Boyajian et al. \citet{Boyajian2012}) have used a mass-radius relation to calculate the radii; Muirhead et al. (\citet{Muirhead2012a}, \citet{Muirhead2012b}) determined new planet properties for the Kepler Objects of Interest (KOIs)  by utilizing the K-band metallicity and $T_{eff}$ relations given by Rojas-Ayala et al. (\citet{RojasAyala2012}); Johnson et al. 
(\citet{Johnson2012}) estimated the stellar properties of (KOI-254) by combining existing photometric relations. \\
$~~~~~~$ Optical and NIR spectroscopic observations are used for estimating the physical properties of the M dwarfs using prominent atomic (Al, Fe, Mg, Ca, Ti, Na, K, etc.) and molecular (FeH, $H_{2}O$, CO, VO, $CH_{4}$ etc.) absorption lines and different spectral indices, which could be obtained with ground-based 2-m class telescope using moderate resolution spectrographs. With the identification and development of empirical tracers which are sensitive to stellar parameters, the idea to establish calibration relations using stars as calibrators, whose parameters are measured directly, is applied extensively by Mann et al. (\citet{Mann2013b}) and Newton et al. (\citet{Newton2015}) to estimate effective temperatures, radii, and luminosities without any help of parallaxes or theoretical stellar models for medium resolution NIR spectra. Rojas-Ayala et al. (\citet{RojasAyala2012}) defined the K-band $H_{2}O$ index and used it to calibrate the index as a spectral type and $T_{eff}$ indicator and also using the index along with the Ks-band Na I doublet and Ca I triplet to develop K-band metallicity estimates for M dwarfs. Using empirical parameter values acquired through interferometric measurements for relatively near-by bright M dwarf calibrators we can create empirical relations applicable to more distant faint stars for which distances or angular diameter are unavailable (Mann et al. \citet{Mann2013b}). Such data can also test models of M dwarf interiors and atmospheres (Boyajian et al. \citet{Boyajian2015}). These fundamental parameters will help to establish a road map for future target selection for transit searches around the young and old population of M dwarfs.  \\
$~~~~~$ To cover the age distribution of M dwarfs, we have prepared a sample of young M-dwarfs from young moving groups and an older population from the galactic field. Over two decades, a variety of young ($\textless$100 Myrs) moving groups have been discovered (e.g. TW Hydra, Pictoris, AB Doradus, Carina, etc.), with distances ($\sim$100 pc) much closer than any star-forming regions (Riedel et al. \citet{Riedel2014}). Pre-main sequence stars are systematically larger than their older counterparts, and would have cooler $T_{eff}$ and larger radii for a given $M_{K}$ magnitude. Using the \href{http://www.tifr.res.in/~daa/tirspec/}{TIRSPEC} instrument on the 2-m \href{https://www.iiap.res.in/iao/cycle.html}{HCT}, we  are  characterizing a sample of M dwarfs using a range of atomic and molecular indices. In the literature, few calibration relations exist either at K-band or H-band spectral coverage at medium resolution (R $\sim$ 2000 to 2700). The cross-disperser mode of TIRSPEC offers simultaneous observations of H- and K-band spectra, and provide more atomic and molecular indices for exploring the better characterization of these cool dwarfs. Furthermore, TIRSPEC spectra offer the resolution R $\sim$1200 which is required to validate the existing relations (which have been obtained at higher resolution).  \\
$~~~~~$ In this paper we present empirical relations between $T_{eff}$, radius, luminosity, spectral type, absolute $K_{s}$ magnitude, metallicity and mass for nearby M dwarf stars, using EWs of several prominent spectral features and $H_{2}O$ indices in the H- and K-bands.

\begin{table*} 
 \centering 
 \caption{Log of HK-band spectroscopic observations using 2-m HCT and observational properties of M Dwarf}
\hspace*{-0.7cm}
 \begin{tabular}{lcccccccc}
  \hline \hline
 Name & R.A.$^{\star}$& Decl.$^{\star}$ & Date of &  2MASS K$^{\dagger}$ & Exp. time (s) & \underline{SNR} & Star & Weather  \\
 & (hh:mm:ss)&(dd:mm:ss)&Observations&(mag) & $\times$ frame no. &H-Band \hspace*{0.1cm} K-Band& &Conditions\\
 \hline
HD 232979 $^{a,b}$&04:37:40.93&+52:53:37.01&12.12.2017&5.047&100$\times$14&73 \hspace*{0.7cm} 53&Flare Star$^{1}$&Partly Clear \\
HD 79210 $^{b,c,d,e}$&09:14:22.77&+52:41:11.79&12.12.2017&3.990&100$\times$10&46 \hspace*{0.7cm} 44&Flare Star$^{1}$& ,, \\
GJ 338B $^{a,b,e}$&09:14:24.68&+52:41:10.91&17.01.2018&4.140&100$\times$10&77 \hspace*{0.7cm} 57&Flare Star$^{1}$&Thin patchy clouds \\
HD 28343 $^{d}$&04:29:00.12&+21:55:21.72&12.12.2017&4.875&100$\times$10&65 \hspace*{0.7cm} 71&Flare Star$^{1}$&Partly Clear \\
GJ 3395 $^{c}$&06:31:01.16&+50:02:48.25&17.02.2016& 7.063&500$\times$6&157 \hspace*{0.7cm} 94&Flare Star$^{1}$&Clear Sky \\
HD 233153 $^{a,b,d}$&05:41:30.73&+53:29:23.29&21.12.2016&5.759&100$\times$18&73 \hspace*{0.7cm} 41&High proper-motion Star$^{2}$& ,, \\
GJ 514 $^{c}$&13:29:59.79&+10:22:37.78&27.02.2017&5.036&100$\times$14&83 \hspace*{0.7cm} 52&High proper-motion Star$^{2}$& ,, \\
GJ 525 &13:45:05.08&+17:47:07.56&12.12.2017&6.220&100$\times$18&34 \hspace*{0.7cm} 32&High proper-motion Star$^{2}$&Partly Clear \\
HD 36395 $^{a,b,e}$&05:31:27.39&- 03:40:38.02&12.12.2017&3.900&100$\times$14&14 \hspace*{0.7cm} 11&High proper-motion Star$^{2}$& ,, \\
GJ 625 $^{c}$&16:25:24.62&+54:18:14.76&27.06.2017&5.833&100$\times$12&83 \hspace*{0.7cm} 57&High proper-motion Star$^{2}$&Thin passing clouds \\
HD 95735 $^{a,b,c,e}$&11:03:20.19&+35:58:11.56&17.02.2016&3.340&100$\times$10&213 \hspace*{0.7cm} 60&Flare Star$^{1}$&Clear Sky \\
HD 115953 &13:19:45.65&+47:46:40.95&17.02.2016&4.494&100$\times$10&240 \hspace*{0.7cm} 80&Double or Multiple Star$^{3}$& ,, \\
V* GX And $^{a,b,e}$&00:18:22.88&+44:01:22.64&21.12.2016&4.020&100$\times$10&252 \hspace*{0.6cm} 198&Flare Star$^{1}$& ,, \\
V* BR Psc &23:49:12.52&+02:24:04.40&22.12.2016&5.043&100$\times$14&158 \hspace*{0.7cm} 95&Variable of BY Dra type$^{4}$&Thin Clouds \\
HD 50281B $^{d}$&06:52:18.04&- 05:11:24.04&12.12.2017& 5.723&100$\times$14&56 \hspace*{0.7cm} 46&High proper-motion Star$^{2}$&Partly Clear \\
GJ 494 $^{c}$&13:00:46.56&+12:22:32.71&27.06.2017& 5.578&100$\times$10&67 \hspace*{0.7cm} 75&Flare Star$^{1}$&Thin passing clouds \\
GJ 649 &16:58:08.84&+25:44:38.97&27.06.2017&5.624&100$\times$12&47 \hspace*{0.7cm} 42&High proper-motion Star$^{2}$& ,, \\
GJ 436 $^{a,b,e}$&11:42:11.09&+26:42:23.65&22.12.2016&6.073&100$\times$12&89 \hspace*{0.7cm} 76&High proper-motion Star$^{2}$&Clear Sky \\
GJ 176 &04:42:55.77&+18:57:29.39&21.12.2016&5.607&100$\times$14&68 \hspace*{0.7cm} 40&High proper-motion Star$^{2}$& ,, \\
GJ 797 B $^{d}$&20:40:44.51&+19:54:03.12&21.12.2016&7.416&500$\times$8&126 \hspace*{0.7cm} 113&Double or Multiple star$^{5}$& ,, \\
HD 119850 $^{a,b,c,d,e}$&13:45:43.77&+14:53;29.47&17.02.2016&4.415&100$\times$10&189 \hspace*{0.7cm} 83&Flare Star$^{1}$& ,, \\
GJ 628 &16:30:18.05&- 12:39:45.32&01.03.2017&5.075&100$\times$10&124 \hspace*{0.7cm} 101&Variable of BY Dra type$^{4}$& ,, \\
GJ 251 $^{d}$&06:54:48.95&+33:16:05.43&27.02.2017&5.275&100$\times$14&95 \hspace*{0.7cm} 75&High proper-motion Star$^{2}$& ,, \\
GJ 109 &02:44:15.51&+25:31:24.11&01.03.2017&5.961&100$\times$14&55 \hspace*{0.7cm} 46&Flare Star$^{1}$&Thin Clouds \\
GJ 48 $^{c}$&01:02:32.23&+71:40:47.33&22.12.2016&5.449&100$\times$12&87 \hspace*{0.7cm} 66&Flare Star$^{1}$&Light wind, Clear \\
GJ 581 $^{e}$&15:19:26.82&- 07:43:20.18&18.01.2018&5.837&100$\times$14&67 \hspace*{0.7cm} 64&Variable of BY Dra type$^{4}$&Clear Sky \\
GJ 661 &17:12:07.91&+45:39:57.21&19.01.2018&4.834&100$\times$10&92 \hspace*{0.7cm} 85&Double or Multiple star$^{5}$& ,, \\
GJ 3378 $^{a,b}$&06:01:11.04&+59:35:49.88&17.02.2016&6.639&500$\times$6&170 \hspace*{0.7cm} 97&High proper-motion Star$^{2}$& ,, \\
GJ 169.1 A &04:31:11.50&+58:58:37.42&17.02.2016&5.717&100$\times$18&160 \hspace*{0.7cm} 94&High proper-motion Star$^{2}$& ,, \\
GJ 273 &07:27:24.49&+05:13:32.83&21.12.2016&4.857&100$\times$10&75 \hspace*{0.7cm} 46&High proper-motion Star$^{2}$& ,, \\
GJ 3522 $^{c}$&08:58:56.32&+08:28:26.06&22.12.2016&5.688&100$\times$12&149 \hspace*{0.7cm} 86&Flare Star$^{1}$& ,, \\
GJ 3379 $^{a,b,e}$&06:00:03.50&+02:42:23.59&22.12.2016&6.042&100$\times$16&129 \hspace*{0.7cm} 83&Flare Star$^{1}$& ,, \\
GJ 447 $^{a,b,e}$&11:47:44.39&+00:48:16.39&18.01.2018&5.654&100$\times$18&105 \hspace*{0.7cm} 82&Flare Star$^{1}$& ,, \\
G 246-33 &03:19:28.87&+61:56:03.79&27.02.2017&8.656&500$\times$10&17 \hspace*{0.7cm} 13&High proper-motion Star$^{2}$& ,, \\
GJ 213 &05:42:09.26&+12:29:21.62&01.03.2017&6.389&100$\times$14&50 \hspace*{0.7cm} 42&Variable of BY Dra type$^{4}$&Thin Clouds \\
GJ 445 $^{c}$&11:47:41.38&+78:41:28.17&01.03.2017&5.954&100$\times$14&101 \hspace*{0.7cm} 82&High proper-motion Star$^{2}$& ,, \\
GJ 402 $^{d}$&10:50:52.03&+06:48:29.26&27.02.2017&6.371&500$\times$10&77 \hspace*{0.7cm} 48&Variable of BY Dra type$^{4}$&Clear Sky \\
GJ 408 &11:00:04.25&+22:49:58.64&12.12.2017&5.503&100$\times$18&41 \hspace*{0.7cm} 33&High proper-motion Star$^{2}$&Partly Clear \\
GJ 277 B &07:31:57.32&+36:13:47.35&17.02.2016&6.755&500$\times$6&168 \hspace*{0.7cm} 53&Variable of BY Dra type$^{4}$&Clear Sky \\
GJ 873 &22:46:49.73&+44:20:02.37&21.12.2016&5.299&100$\times$16&118 \hspace*{0.7cm} 99&Flare Star$^{1}$& ,, \\
GJ 1156 $^{c}$&12:18:59.39&+11:07:33.77&19.01.2018&7.570&500$\times$10&12 \hspace*{0.7cm} 10&Flare Star$^{1}$& ,, \\
GJ 3454 &07:36:25.11&+07:04:43.13&19.01.2018&7.282&500$\times$10&87 \hspace*{0.7cm} 85&Flare Star$^{1}$& ,, \\
GJ 166 C $^{c,d,e}$&04:15:21.73&- 07:39:17.36&19.01.2018&5.962&100$\times$16&92 \hspace*{0.7cm} 82&Flare Star$^{1}$& ,, \\
GJ 3304 &04:38:12.55&+28:13:00.11&01.03.2017&7.326&500$\times$9&42 \hspace*{0.7cm} 34&Flare Star$^{1}$&Thin Clouds \\
GJ 324 B $^{d}$&08:52:40.86&+28:18:58.82&22.12.2016&7.666&500$\times$11&73 \hspace*{0.7cm} 57&High proper-motion Star$^{2}$&Clear Sky \\
GJ 3421 &07:03:55.65&+52:42:07.60&18.01.2018&7.776&500$\times$10&79 \hspace*{0.7cm} 75&High proper-motion Star$^{2}$& ,, \\
GJ 905 $^{a,b,c,d,e}$&23:41:54.98&+44:10:40.78&19.01.2018&5.929&100$\times$16&97 \hspace*{0.7cm} 89&Variable of BY Dra type$^{4}$& ,, \\
V* CU Cnc &08:31:37.57&+19:23:39.39&22.12.2016&6.603&500$\times$7&114 \hspace*{0.7cm} 68&Eclipsing Binary$^{6}$& ,, \\
GJ 1286 &23:35:10.50&- 02:23:21.44&22.12.2016&8.183&500$\times$6&27 \hspace*{0.7cm} 20&High proper-motion Star$^{2}$&Light wind, Clear \\
GJ 473 $^{b}$&12:33:17.38&+09:01:15.77&01.03.2017&6.042&100$\times$14&128 \hspace*{0.7cm} 114&Flare Star$^{1}$&Clear Sky \\
GJ 406 $^{c,d}$&10:56:28.86&+07:00:52.77&18.01.2018&6.084&500$\times$10&45 \hspace*{0.7cm} 42&Flare Star$^{1}$& ,, \\
GJ 1083 $^{c}$&05:40:25.70&+24:48:09.02&19.01.2018&8.039&500$\times$10&48 \hspace*{0.7cm} 44&Flare Star$^{1}$& ,, \\
GAT 1370 $^{b,c}$&02:53:00.89&+16:52:52.64&22.12.2016&7.585&500$\times$10&10 \hspace*{0.7cm} 8&Low-mass star$^{7}$& ,, \\
  
  \hline
\end{tabular}
 \vspace{1ex} 
 
\hspace*{-17.cm} \textbf{Notes.}
 
          \raggedright \small $^{a}$ M-dwarf $T_{eff}$, Radius $\&$ $L_{bol}$ calibration samples (see Table \ref{tab:table4_teff_rad_lum_cal}). \\
          \raggedright \small $^{b}$ M-dwarf spectral type calibration samples (see Table \ref{tab:table5_sptype_cal}). \\
          \raggedright \small $^{c}$ M-dwarf M$_{k}$ calibration samples (see Table \ref{tab:table6_mag_cal}). \\
          \raggedright \small $^{d}$ M-dwarf metallicity calibration samples (see Table \ref{tab:table7_metallicity_cal}). \\
          \raggedright \small $^{e}$ M-dwarf mass calibration samples (see Table \ref{tab:table8_mass_cal}).  \\
          \raggedright \small $^{\star}$ Positions are taken from International Celestial Reference System \href{https://www.iers.org/IERS/EN/Science/ICRS/ICRS.html}{(ICRS)}.\\
          \raggedright \small $^{\dagger}$ Apparent $K_{s}$ magnitudes from 2MASS All Sky catalogue. \\
           \textbf{References for star properties.} \small $^{1}$ [GKL99] (Gershberg et al. \citet{GERSHBERG1999});  $^{2}$ Frith, J. (\citet{Frith2013});  $^{3}$ Tokovinin, A. A. (\citet{Tokovinin1997});  $^{4}$ Samus, N. N. et al. (\citet{Samus2002});  $^{5}$ TERRIEN R.C. et al. \citet{Terrien2015};  $^{6}$ LOPEZ-MORALES M. \citet{LopezMorales2007};  $^{7}$ GAGNE J. et al. \citet{Gagne2015}.\\
 \label{tab:table1_obslog}

\end{table*}
 
\section{Sample Selection and Observation :}

We observed a sample of 53 M dwarfs in total with declinations higher than -$15^{\circ}$. The sample includes a variety of subsets that act as calibrators by providing a range of empirical properties (see below in Table \ref{tab:table1_obslog}). The observations were performed using the medium resolution TIRSPEC instrument on the 2-m HCT at Hanle, India over 10 nights during different observing cycles from 2016-2018. Due to limited observing nights we preferably tried to select nearby bright (K-mag within 8.0) objects so that we could observe using 100s exposure in multiple frames to cover the spectral ranges up to M7V. Our observational samples are:

\begin{figure*}
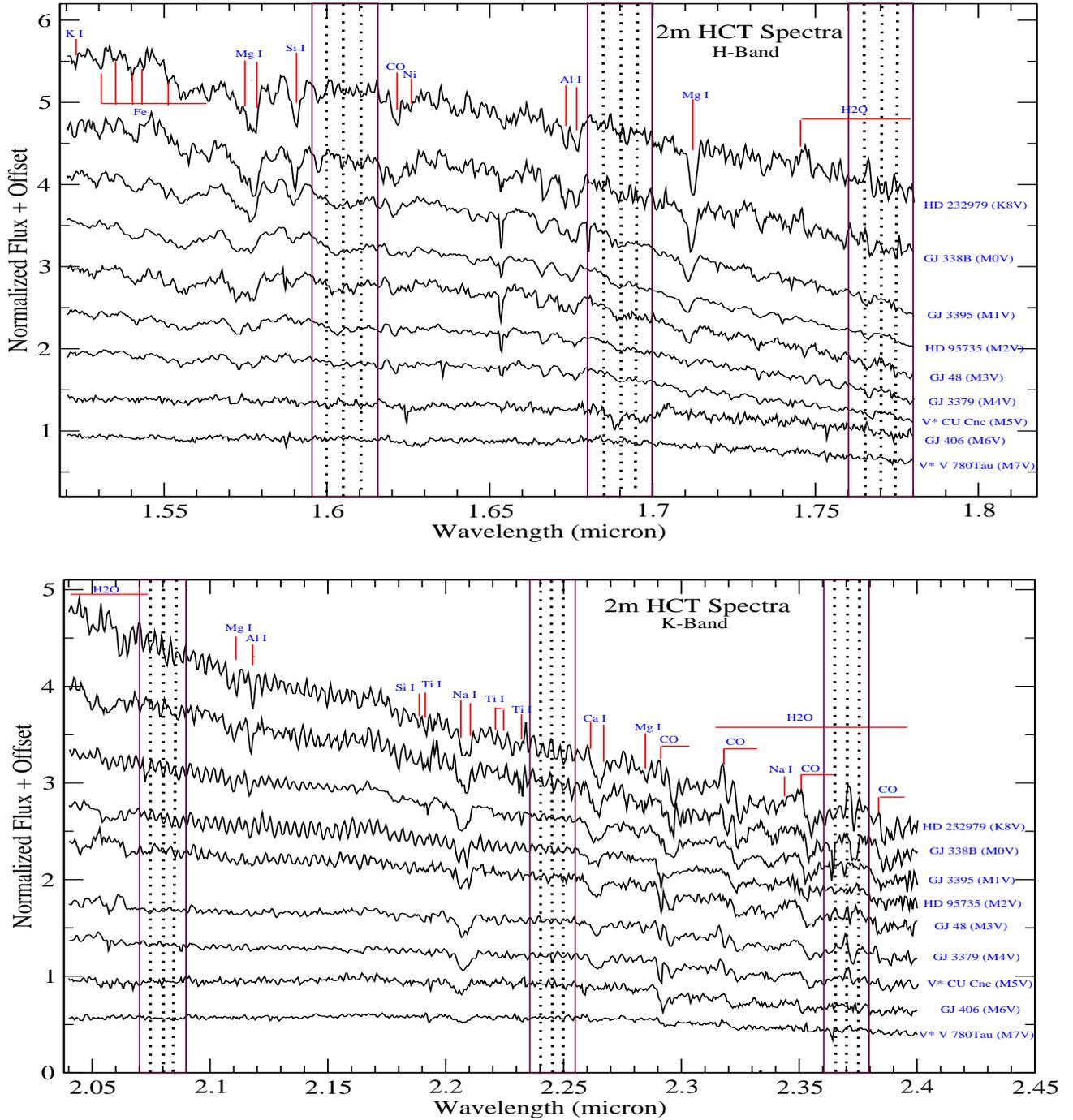

\centering
\begin{tabular}{cc}
\hspace*{-0.25cm} \includegraphics[width= 6.6 in, height = 3.5 in, clip]{fig1_allspec_h1.eps} \\[2\tabcolsep]
\hspace*{0.15cm} \includegraphics[width= 6.7 in, height = 3.5 in, clip]{fig1_allspec_k1.eps}
\end{tabular}
\caption{A few example spectra of H- and K-Band with the sequence of the spectral types from K8V to M7V observed using the 2-m HCT are shown. Some of the most prominent and significant atomic and molecular absorption features are indicated. The portions dominated by water absorption and almost free from any features from (1.595-1.780) $\mu$m in H-band and from (2.07-2.38) $\mu$m in K-band are shown in boxes filled with dotted lines.} 
\label{fig1}
\end{figure*}

\begin{table*} 
 \centering 
 \caption{Estimation of EWs and $H_{2}O$ indices in H- and K-band for the TIRSPEC M-Dwarf samples :}
 \begin{tabular}{lccccccc}
  \hline \hline
 Name & EW ( Mg I )& EW ( Al-a ) & EW ( Mg I ) &  $H_{2}O$-H & EW ( Na I ) & EW ( Ca I )  & $H_{2}O$-K  \\
 & (1.57 $\mu$m)&(1.67 $\mu$m)&(1.71 $\mu$m)& & (2.21 $\mu$m) &(2.26 $\mu$m) &\\
 \hline
HD 232979 &3.979$\pm$0.358&0.848$\pm$0.062&2.495$\pm$0.377&0.940$\pm$0.004&3.913$\pm$0.240&3.364$\pm$0.292&1.074$\pm$0.009 \\
HD 79210 &3.553$\pm$0.388&0.651$\pm$0.113&3.323$\pm$0.374&0.970$\pm$0.008&5.206$\pm$0.386&3.414$\pm$0.364&1.044$\pm$0.005 \\
GJ 338B &4.409$\pm$0.164&1.119$\pm$0.226&2.449$\pm$0.261&0.909$\pm$0.008&4.991$\pm$0.389&3.704$\pm$0.181&0.992$\pm$0.005 \\
HD 28343 &5.548$\pm$0.284&1.407$\pm$0.131&2.659$\pm$0.323&0.936$\pm$0.008&5.893$\pm$0.490&5.184$\pm$0.534&1.060$\pm$0.009 \\
GJ 3395 &4.262$\pm$0.550&0.894$\pm$0.036&1.532$\pm$0.246&0.900$\pm$0.003&5.265$\pm$0.276&3.821$\pm$0.225&0.961$\pm$0.004 \\
HD 233153 &3.871$\pm$0.321&1.196$\pm$0.039&1.476$\pm$0.475&0.949$\pm$0.007&5.382$\pm$0.696&3.871$\pm$0.629&0.959$\pm$0.017 \\
GJ 514 &2.770$\pm$0.087&2.854$\pm$0.112&2.303$\pm$0.509&0.928$\pm$0.009&5.242$\pm$0.533&3.644$\pm$0.247&1.000$\pm$0.009 \\
GJ 525 &2.594$\pm$0.832&0.991$\pm$0.235&2.617$\pm$0.297&0.931$\pm$0.006&3.990$\pm$0.501&1.452$\pm$0.551&0.944$\pm$0.032 \\
HD 36395 &3.565$\pm$0.764&0.586$\pm$0.226&1.003$\pm$0.580&0.909$\pm$0.012&0.477$\pm$0.0&7.037$\pm$0.969&0.898$\pm$0.039 \\
GJ 625 &1.450$\pm$0.396&0.945$\pm$0.085&0.931$\pm$0.222&0.901$\pm$0.007&5.322$\pm$0.309&2.955$\pm$0.393&0.874$\pm$0.006 \\
HD 95735 &2.661$\pm$0.155&0.726$\pm$0.019&0.333$\pm$0.127&0.897$\pm$0.003&3.717$\pm$0.748&2.057$\pm$0.338&0.930$\pm$0.006 \\
HD 115953 &3.555$\pm$0.203&0.920$\pm$0.049&0.973$\pm$0.092&0.923$\pm$0.004&4.745$\pm$0.345&3.571$\pm$0.285&0.977$\pm$0.002 \\
V* GX And &2.300$\pm$0.066&0.860$\pm$0.046&1.250$\pm$0.141&0.918$\pm$0.002&3.640$\pm$0.371&2.532$\pm$0.152&0.949$\pm$0.004 \\
V* BR Psc &2.986$\pm$0.091&0.861$\pm$0.041&0.658$\pm$0.112&0.867$\pm$0.008&2.720$\pm$0.469&2.660$\pm$0.447&0.943$\pm$0.006 \\
HD 50281B &2.520$\pm$0.730&0.935$\pm$0.112&1.634$\pm$0.293&0.902$\pm$0.007&5.581$\pm$0.475&4.488$\pm$0.380&0.954$\pm$0.019 \\
GJ 494 &3.373$\pm$0.188&1.586$\pm$0.089&1.527$\pm$0.266&0.887$\pm$0.009&6.117$\pm$0.374&5.218$\pm$0.119&0.914$\pm$0.009 \\
GJ 649 &2.849$\pm$0.148&1.075$\pm$0.088&1.312$\pm$0.309&0.958$\pm$0.007&5.372$\pm$0.727&3.736$\pm$0.444&0.968$\pm$0.007 \\
GJ 436 &3.014$\pm$0.153&0.853$\pm$0.043&1.189$\pm$0.159&0.877$\pm$0.002&4.873$\pm$0.294&3.524$\pm$0.463&0.942$\pm$0.009 \\
GJ 176 &2.195$\pm$0.256&1.307$\pm$0.063&0.606$\pm$0.244&0.942$\pm$0.008&5.357$\pm$0.363&3.065$\pm$0.610&0.932$\pm$0.020 \\
GJ 797 B &1.682$\pm$0.566&2.365$\pm$0.134&1.292$\pm$0.161&0.913$\pm$0.007&4.126$\pm$0.864&2.904$\pm$0.130&0.943$\pm$0.005 \\
HD 119850 &3.522$\pm$0.186&0.966$\pm$0.046&1.153$\pm$0.168&0.911$\pm$0.002&4.018$\pm$0.504&2.870$\pm$0.160&0.950$\pm$0.005 \\
GJ 628 &1.608$\pm$0.204&0.969$\pm$0.047&0.666$\pm$0.130&0.846$\pm$0.006&5.001$\pm$0.197&3.021$\pm$0.222&0.830$\pm$0.007 \\
GJ 251 &0.996$\pm$0.199&1.615$\pm$0.075&0.991$\pm$0.070&0.920$\pm$0.008&4.800$\pm$0.257&2.764$\pm$0.222&0.946$\pm$0.006 \\
GJ 109 &0.570$\pm$0.263&0.776$\pm$0.063&0.735$\pm$0.293&0.843$\pm$0.004&4.509$\pm$0.441&2.687$\pm$0.328&0.876$\pm$0.004 \\
GJ 48 &3.728$\pm$0.455&1.173$\pm$0.083&1.542$\pm$0.098&0.861$\pm$0.008&4.120$\pm$0.408&4.014$\pm$0.318&0.903$\pm$0.008 \\
GJ 581 &2.006$\pm$0.259&0.450$\pm$0.094&0.120$\pm$0.085&0.884$\pm$0.009&3.923$\pm$0.585&3.529$\pm$0.320&0.847$\pm$0.009 \\
GJ 661 &1.926$\pm$0.155&0.947$\pm$0.118&0.356$\pm$0.144&0.873$\pm$0.005&2.529$\pm$0.200&3.005$\pm$0.368&0.869$\pm$0.005 \\
GJ 3378 &1.780$\pm$0.272&0.548$\pm$0.044&0.489$\pm$0.123&0.856$\pm$0.007&5.146$\pm$0.307&2.877$\pm$0.222&0.883$\pm$0.006 \\
GJ 169.1 A &1.769$\pm$0.090&0.710$\pm$0.050&0.475$\pm$0.104&0.873$\pm$0.007&7.209$\pm$0.258&3.164$\pm$0.080&0.888$\pm$0.006 \\
GJ 273 &1.844$\pm$0.100&0.813$\pm$0.030&0.424$\pm$0.087&0.864$\pm$0.009&4.202$\pm$0.637&2.578$\pm$0.833&0.851$\pm$0.008 \\
GJ 3522 &1.806$\pm$0.168&0.479$\pm$0.057&0.261$\pm$0.090&0.811$\pm$0.004&4.771$\pm$0.290&2.764$\pm$0.167&0.864$\pm$0.004 \\
GJ 3379 &1.883$\pm$0.125&0.465$\pm$0.058&0.395$\pm$0.077&0.827$\pm$0.008&5.937$\pm$0.235&2.516$\pm$0.284&0.856$\pm$0.006 \\
GJ 447 &0.936$\pm$0.160&0.502$\pm$0.093&0.503$\pm$0.096&0.823$\pm$0.009&5.535$\pm$0.216&2.501$\pm$0.222&0.829$\pm$0.002 \\
G 246-33 &  \textemdash &\textemdash &\textemdash &0.901$\pm$0.014&2.999$\pm$0.944& \textemdash &0.713$\pm$0.032 \\
GJ 213 &0.459$\pm$0.212&0.861$\pm$0.223&0.240$\pm$0.188&0.826$\pm$0.008&2.963$\pm$0.936&2.682$\pm$0.370&0.912$\pm$0.021 \\
GJ 445 &1.134$\pm$0.206&0.740$\pm$0.061& \textemdash &0.829$\pm$0.005&3.272$\pm$0.118&1.828$\pm$0.329&0.828$\pm$0.003 \\
GJ 402 &0.488$\pm$0.122&2.422$\pm$0.096&0.638$\pm$0.186&0.874$\pm$0.006&7.089$\pm$0.371&2.342$\pm$0.244&0.894$\pm$0.005 \\
GJ 408 &1.149$\pm$0.321&0.374$\pm$0.124&0.919$\pm$0.261&0.901$\pm$0.005&7.689$\pm$0.817&3.846$\pm$0.676&0.903$\pm$0.007 \\
GJ 277 B &2.577$\pm$0.082&0.722$\pm$0.076&0.645$\pm$0.064&0.850$\pm$0.008&5.709$\pm$0.387& 3.879$\pm$0.270&0.872$\pm$0.003 \\
GJ 873 &0.596$\pm$0.127&2.198$\pm$0.034&0.912$\pm$0.138&0.882$\pm$0.009&5.469$\pm$0.138&3.222$\pm$0.166&0.906$\pm$0.008 \\
GJ 1156 &1.216$\pm$0.605& \textemdash & \textemdash &0.942$\pm$0.047&5.175$\pm$0.964& \textemdash &0.870$\pm$0.055 \\
GJ 3454 &0.385$\pm$0.170&0.985$\pm$0.179&0.372$\pm$0.123&0.803$\pm$0.003&5.785$\pm$0.601&2.758$\pm$0.443&0.809$\pm$0.002 \\
GJ 166 C &1.367$\pm$0.259&0.664$\pm$0.066& \textemdash &0.798$\pm$0.006&3.885$\pm$0.217&2.467$\pm$0.294&0.798$\pm$0.007 \\
GJ 3304 &  \textemdash &0.539$\pm$0.101&4.271$\pm$0.988&0.812$\pm$0.005&4.599$\pm$0.641&1.726$\pm$0.777&0.844$\pm$0.014 \\
GJ 324 B &2.380$\pm$0.360& \textemdash &0.302$\pm$0.093&0.852$\pm$0.008&8.635$\pm$0.650&3.331$\pm$0.266&0.908$\pm$0.004 \\
GJ 3421 &0.151$\pm$0.0&0.469$\pm$0.089&0.176$\pm$0.0&0.746$\pm$0.009&3.143$\pm$0.426&1.119$\pm$0.216&0.802$\pm$0.004 \\
GJ 905 &1.122$\pm$0.220&0.630$\pm$0.033& \textemdash &0.847$\pm$0.005&6.857$\pm$0.404&2.406$\pm$0.155&0.786$\pm$0.004 \\
V* CU Cnc &2.281$\pm$0.320&0.764$\pm$0.074&0.826$\pm$0.091&0.827$\pm$0.005&7.413$\pm$0.370&4.531$\pm$0.352&0.875$\pm$0.006 \\
GJ 1286 &3.590$\pm$0.660& \textemdash &0.820$\pm$0.443&0.783$\pm$0.015&5.634$\pm$0.0&1.832$\pm$0.0&0.832$\pm$0.024 \\
GJ 473 &  \textemdash &0.420$\pm$0.095& \textemdash &0.809$\pm$0.006&5.564$\pm$0.108&2.395$\pm$0.371&0.790$\pm$0.006 \\
GJ 406 &  \textemdash  &0.179$\pm$0.067&0.282$\pm$0.0&0.918$\pm$0.012&3.271$\pm$1.033&0.801$\pm$0.338&0.771$\pm$0.008 \\
GJ 1083 &0.116$\pm$0.0            &0.197$\pm$0.063&0.182$\pm$0.0&0.778$\pm$0.008&4.582$\pm$0.208&2.958$\pm$0.656&0.763$\pm$0.007 \\
GAT 1370 &6.071$\pm$2.130 &\textemdash &3.907$\pm$1.637&0.603$\pm$0.015&10.577$\pm$0.0 & \textemdash  &0.677$\pm$0.027 \\

  \hline
\end{tabular}
 \label{tab:table2_ew_waterindices_cal}

\end{table*}

\begin{figure*}
\centering
\begin{tabular}{c}
\hspace*{-0.85cm} \includegraphics[width= 5.3 in, height = 7.5 in, angle=-90, clip]{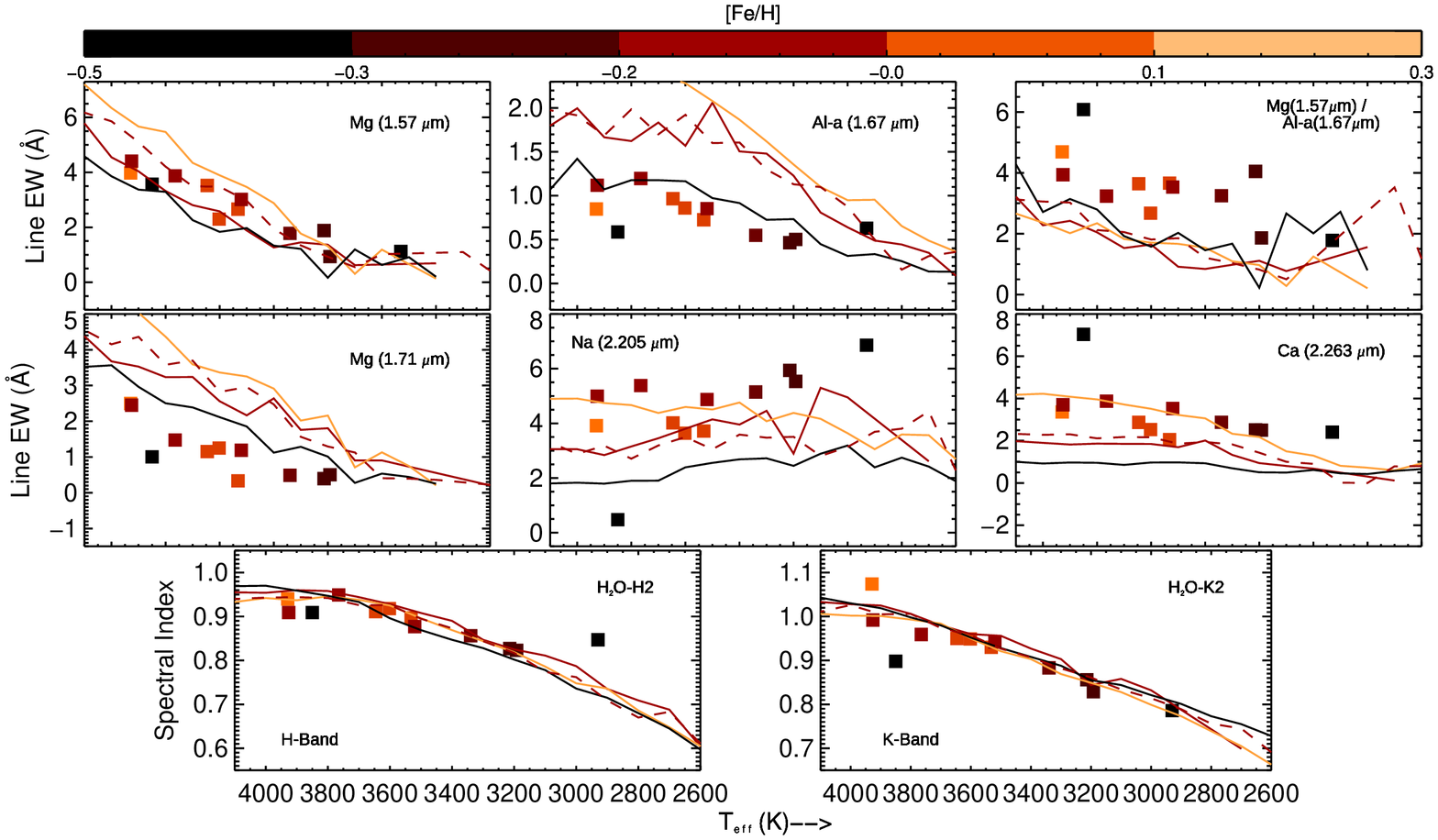}
\end{tabular}
\caption{EWs, EW ratio, and $H_{2}O$ indices at H and K-band are plotted against the measured $T_{eff}$ for our observed calibration sample. We used the value of $T_{eff}$ and [Fe/H] estimated in this work. We show the overplot of derived EWs, EW ratios and water indices from BT-Settl model spectra. The red solid line is for log g = 5.0 and the red dashed line is for log g = 4.5, both are for solar metallicity model [Fe/H = 0]. For log g = 5.0 we plot for two other metallicities [Fe/H], -0.5dex (black solid) and +0.5dex (orange solid).}
\label{fig:fig2}
\end{figure*}

\begin{enumerate}
\item[1.] \textsc{Calibrators with interferometrically measured parameters:} Among our observed sample we used a subset of 12 M dwarfs to calibrate the $T_{eff}$, radius and luminosity (11 stars). The parameters of 7 stars among the calibrators are taken from Newton et al. \citet{Newton2015}), with the others taken from Mann et al. (\citet{Mann2015}). The stars in our calibration samples have $T_{eff}$ from 2930K$-$3930K, stellar radius from 0.189$R_{\sun}-0.608R_{\sun}$ and bolometric luminosity (log $L/L_{\sun}$ ) from \textendash 1.10 \ to \textendash 2.44.

\item[2.] \textsc{Metallicity Calibrators:} We selected 12 M-dwarfs for [Fe/H] and 9 stars for [M/H] calibration for our work. Our calibrators are in common proper-motion (CPM) pairs where the spectroscopic metallicity measurements for the FGK-primaries are performed by Valenti and Fischer (\citet{Valenti2005}, SPOCS I (Spectroscopic Properties of Cool Stars) catalogue). To enrich our sample with young objects we included spectra of three stars, GJ 406, GJ 402 and GJ 905 with supersolar metallicity values estimated by Mann et al. (\citet{Mann2015}).

\item[3.] \textsc{Objects to calibrate Spectral Type, Mass and $M_{k_{s}}$:} We have created three different subsets from our observed M dwarfs for the calibration of spectral type (sp. type), stellar mass and absolute K magnitude ($M_{k_{s}}$). Among 15 calibrator stars for sp. type, the spectral type of 5 M-dwarfs have been reported by Kirkpatrick et al. (\citet{Kirkpatrick1991}, \citet{Kirkpatrick1995}, \citet{Kirkpatrick1999}) and the rest are selected from the CARMENES input catalogue (Alonso Floriano et al. \citet{AlonsoFloriano2015} and A. Reiners et al., \citet{Reiners2018}). To calibrate absolute $K_{s}$ magnitude we chose 16 calibrators with known 2MASS (Two Micron All Sky Survey, Cutri et al. \citet{Cutri2003}) apparent magnitudes (Cutri et al. \citet{Cutri2003}) and trigonometric parallaxes taken from the Gaia Collaboration et al. (Gaia DR1, \citet{Gaia2016} ; Gaia DR2,  \citet{Gaia2018}), van Leeuwen (\citet{vanLeeuwen2007}), Henry et al. (\citet{Henry2006}), Gatewood at al.\citet{Gatewood2008} and Dittmann et al.\citet{Dittmann2014}. We selected a sample of 12 M dwarfs as calibrators of stellar mass with photometrically estimated mass and luminosity values for 8 objects from Mann et al. \citet{Mann2013a} and 4 objects from Mann et al. (\citet{Mann2015}).

\item[4.] \textsc{Potential Planet Hosts:} We included 10 possible planet-host candidates in our observing sample identified in the NASA exoplanet archive and mentioned by Rojas-Ayala et al. (\citet{RojasAyala2012}). Our objects are within 20 pc and more than half the sample are within 8 pc.
\end{enumerate}

The NIR spectra of our sample of M dwarfs are obtained using the \href{http://www.tifr.res.in/~daa/tirspec/}{TIRSPEC} instrument which has a $1024\times 1024$ Hawaii-1 detector array providing the resolution ($R\approx$ 1200). We selected the $1.97"\times 50"$ slit (S3) and observed in JHKX cross disperse mode in a single exposure that covers the wavelength range from 1.50 - 2.45 \micron. We took 3 dark frames before and after each night's observations to subtract the dark current of the detector from our spectra. To prevent any significant part of the spectra falling on a bad pixel and to reduce noise we performed dithering along the slit using an ABBA slit-nodding pattern. We took special care for binary stars by rotating the slit in such a way that sky spectra was free from the contamination from the another star. To improve the signal to noise ratio (SNR) we calculated the required exposure times and gave exposures of 100s and 500s (for faint stars) in multiple frames. We took Argon lamp spectra and tungsten calibration lamp spectra for calibration purposes and also observed NIR spectroscopic standard stars (A0V/A1V) in nearby airmass range for telluric line removal. A log of our observations is given in Table \ref{tab:table1_obslog} where we mention the estimated SNR for our M dwarfs both in H and K-band.

\section{Data Reduction and SNR calculation:}
The data reduction was performed using standard tasks of the Image Reduction and Analysis Facility ({$IRAF^{\href{http://iraf.noao.edu/}{3}}$}) and a Semi-automated TIRSPEC data reduction pipeline developed by Ninan et al. (\citet{Ninan2014}). In this process first we chose and inspect the list of all science frames and their filters, corresponding flats and Argon lists to reduce the spectra. We median combined the images for the same dither positions and took the average combined images of different dither positions. Spectra was then flat field corrected and sky subtraction was done by differencing the exposures of the two nod positions (A and B). Each sky subtracted exposure was divided by a normalized master flat and the wavelength calibration performed using the Argon lamp spectra. Taking a high-resolution model of Vega, a telluric spectrum was acquired by eliminating the hydrogen lines of the observed A0V star. Normally the A0V star has a NIR spectrum which contains only H I absorption lines contributed from the star's atmosphere with a smooth continuum, free from all other absorption features and metal lines. The M dwarf target spectra were divided by the telluric spectrum and thus largely corrected for telluric contamination from earth's atmosphere. At last the spectra were flux calibrated in H and K using 2MASS photometry. \\
$~~~~~~~$ The SNR of the NIR spectra and the errors associated with the flux measurements were derived using a simple Spectroscopic SNR measurement algorithm \texttt{$DER\_SNR$}\footnote{\url{http://www.stecf.org/software/ASTROsoft/DER_SNR/}}. The basic assumptions taken into account are, the noise is normally distributed as well as uncorrelated in wavelength bins within two pixels and the signal over 5 or more pixels are approximated as a straight line. In general, these conditions are met for most spectra and further details are given in Stoehr et al. \citet{Stoehr2008}.

\begin{table}
 \centering
 \caption{Spectral Features and Continuum Points}
 \begin{tabular}{lccc}
  \hline
  Feature & Feature window & Blue Continuum & Red Continuum \\
   & ($\mu$m) & ($\mu$m) & ($\mu$m)\\
  \hline
  Mg (1.57$\mu$m)$^{a}$ & 1.5738 \hspace*{0.2cm}1.5790 & 1.5640\hspace*{0.2cm} 1.5680 & 1.5790\hspace*{0.2cm} 1.5810\\
  Al-a (1.67$\mu$m)$^{a}$ 	& 1.6715\hspace*{0.2cm} 1.6735	& 1.6550\hspace*{0.2cm} 1.6650	& 1.6780 \hspace*{0.2cm}1.6820	\\
  Mg (1.71$\mu$m)$^{a}$ 	& 1.7085 \hspace*{0.2cm} 1.7145	& 1.7078\hspace*{0.2cm} 1.7108 & 1.7115 \hspace*{0.2cm}1.7175	\\
  Na (2.21$\mu$m)$^{b}$ 	& 2.2040 \hspace*{0.2cm} 2.2110	& 2.1930\hspace*{0.2cm} 2.1970	& 2.2140 \hspace*{0.2cm}2.2200	\\     
  Ca (2.26$\mu$m)$^{b}$ 	& 2.2605\hspace*{0.2cm}  2.2675 	& 2.2558\hspace*{0.2cm} 2.2603	& 2.2678\hspace*{0.2cm} 2.2723	\\
  \hline
 \end{tabular}
 \vspace{1ex} \\
   \raggedright  \texttt{Wavelengths are given in vacuum}.\\
           \raggedright  \texttt{{{$^{a}$ Source: Newton et al. ( \citet{Newton2015})}}}\\
     \raggedright  \texttt{{$^{b}$ Source: Rojas-Ayala et al.\citet{RojasAyala2012}}}  \label{tab:table3_spectralfeature}

\end{table} 

\begin{figure*}
\centering
\begin{tabular}{ccc}
\includegraphics[width= 3.2 in, height = 2.5 in, clip]{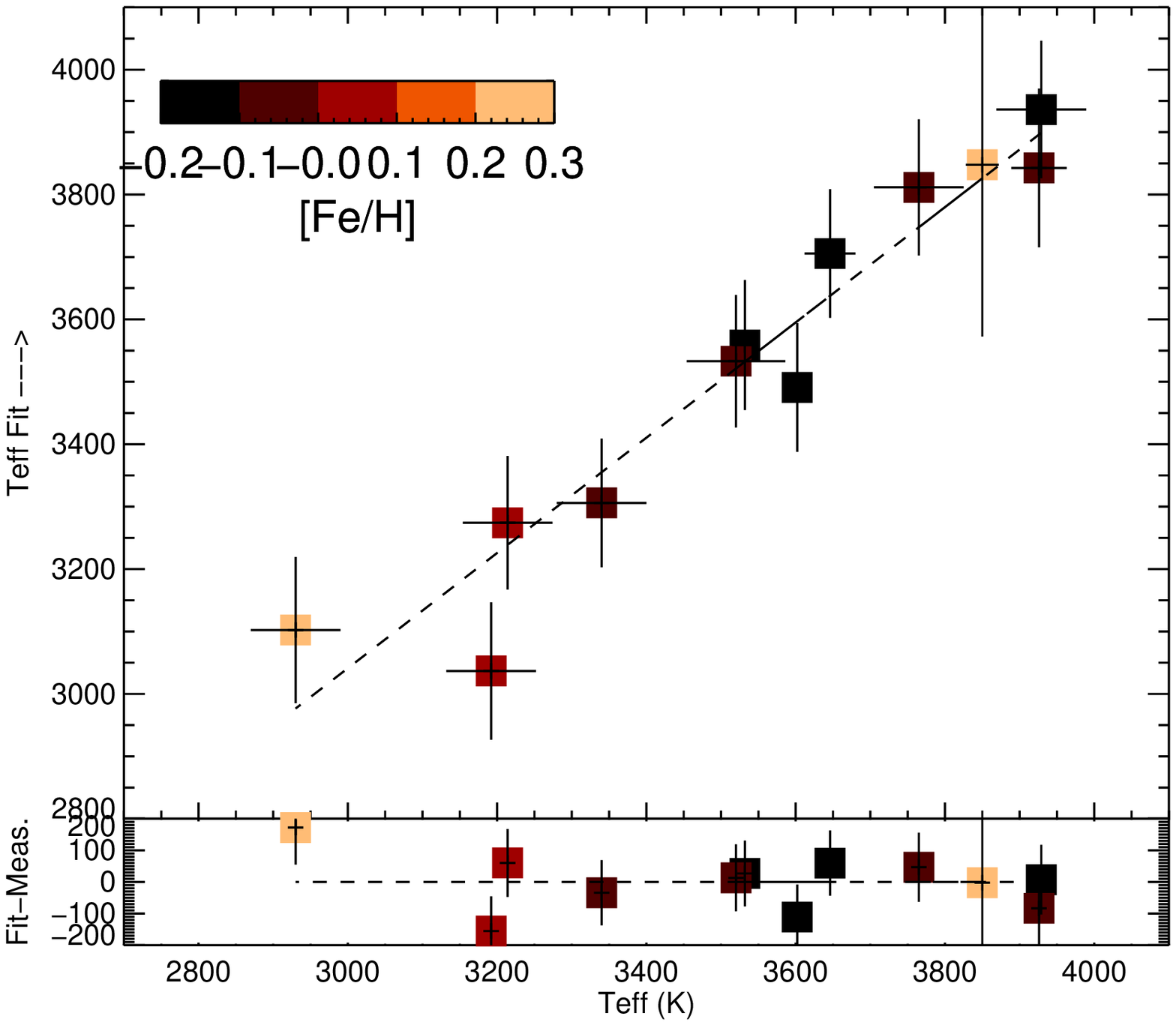} \hspace*{0.7 cm}
\includegraphics[width= 3.2 in, height = 2.5 in, clip]{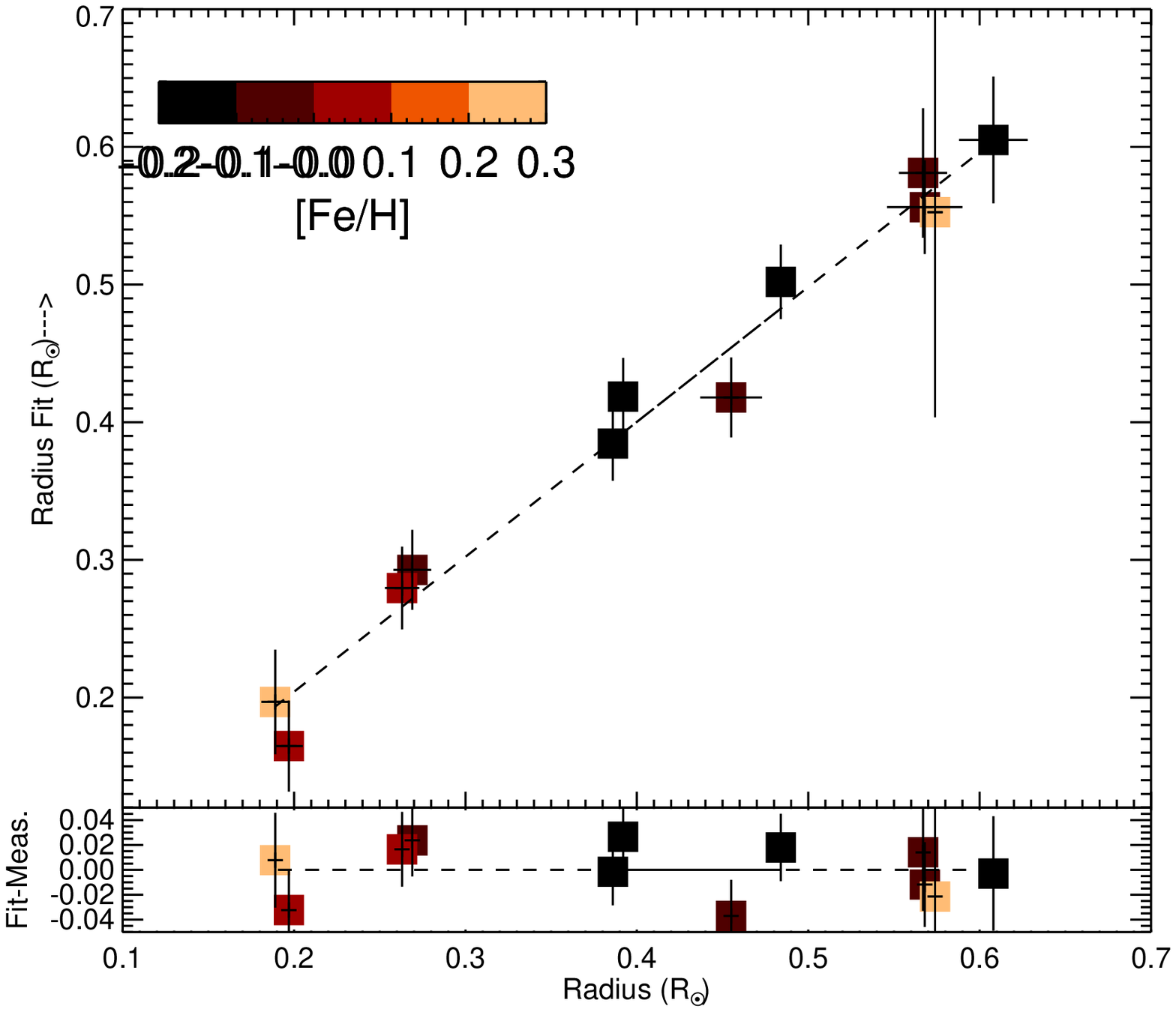}\\[2\tabcolsep]
\includegraphics[width= 3.2 in, height = 2.5 in, clip]{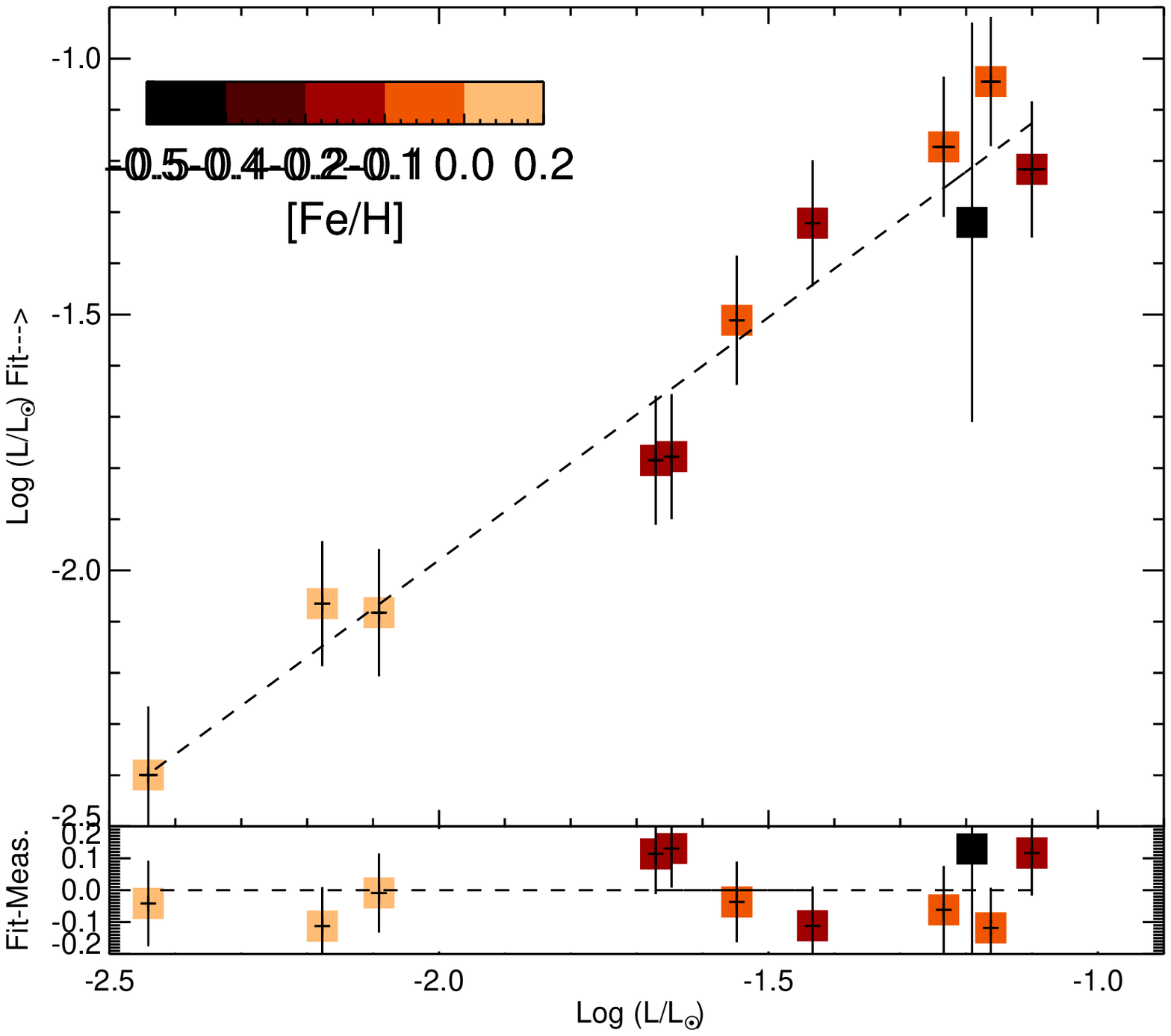}
\end{tabular}
\caption{The plots of our best-fitting calibration relationships for $T_{eff}$ (top left), radius (top right) and Log ($L_{bol}$) (bottom) are shown here. The horizontal axes show the interferometrically measured stellar parameters and some values
taken from Mann et al. (\citet{Mann2015}). In the upper panel of each plot, the vertical axes represent our inferred parameters; in the lower panel, the vertical axes show the residual between our
best-fitting values and the directly measured values. The data points are color coded by the [Fe/H] values calculated in this work.}
\label{fig:fig3}
\end{figure*}

\begin{table*}
 \centering
 \caption{INFERRED $T_{eff}$, RADIUS AND $L_{bol}$ FOR THE CALIBRATION SAMPLES}
  \label{tab:table4_teff_rad_lum_cal}
 \begin{tabular}{lcccccc}
  \hline \hline
\large{Star} &  & \large{Measured Parameters} &  &  & \large{Inferred Parameters} &  \\
   & $T_{eff}$ & Radius & $L_{bol}$ & $T_{eff}$ $^{c}$ & Radius$^{d}$ & $L_{bol}$ $^{e}$  \\
   & (K) & $R_{\sun}$ & log $L/L_{\sun}$  & (K) & $R_{\sun}$ & log $L/L_{\sun}$ \\

 \hline
HD 232979$^{a}$  & 3929$\pm$60 & 0.608$\pm$0.0200 & \textendash 1.1003$\pm$0.0210 & 3936.14$\pm$110 & 0.605$\pm$ 0.046&\textendash 1.216$\pm$	0.133 \\
GJ 338B$^{a}$   & 3926$\pm$37 & 0.567$\pm$0.0137  &  \textendash 1.1627$\pm$0.0145 & 3842.71$\pm$127 &0.581$\pm$ 0.047 &\textendash 1.045$\pm$0.126 \\
HD 233153$^{b}$    & 3765$\pm$60  & 0.568$\pm$0.0220 &  \textendash 1.2340$\pm$0.0170 & 3811.58$\pm$109&0.556$\pm$ 0.034&\textendash 1.172$\pm$0.137 \\
HD 36395$^{a}$  & 3850$\pm$22 & 0.574$\pm$0.0044 &  \textendash 1.1910$\pm$0.0094  &3847.79$\pm$275 &0.553$\pm$ 0.149 &\textendash 1.320$\pm$0.390 \\
HD 95735$^{a}$  & 3532$\pm$17 & 0.392$\pm$0.0033 &  \textendash 1.6710$\pm$0.0061  &3558.98$\pm$104 &0.419$\pm$ 0.028 &\textendash 1.785$\pm$0.126 \\
V* GX And$^{a}$   & 3602$\pm$13 & 0.386$\pm$0.0021 &  \textendash 1.6470$\pm$0.0052  &3490.94$\pm$103 &0.385$\pm$ 0.027 &\textendash 1.778$\pm$0.122 \\
GJ 436$^{a}$   & 3520$\pm$66 & 0.455$\pm$0.0180 & \textendash 1.5480$\pm$0.0110  &3533.07$\pm$106 &0.418$\pm$ 0.029 &\textendash 1.511$\pm$0.126 \\
HD 119850$^{a}$  & 3646$\pm$34 & 0.484$\pm$0.0084 & \textendash 1.4330$\pm$0.0060  &3705.58$\pm$103 &0.502$\pm$ 0.027 &\textendash 1.322$\pm$0.123 \\
GJ 3378$^{b}$   & 3340$\pm$60  & 0.269$\pm$0.0110 &  \textendash 2.0910$\pm$0.0085  &3306.02$\pm$103 &0.293$\pm$ 0.029 &\textendash 2.082$\pm$0.124 \\
GJ 3379$^{b}$     & 3214$\pm$60 & 0.263$\pm$0.0100 & \textendash 2.1770$\pm$0.0120  &3274.17$\pm$107 &0.280$\pm$ 0.030 &\textendash 2.065$\pm$ 0.122 \\  
GJ 447$^{b}$   & 3192$\pm$60 & 0.197$\pm$0.0080 & \textendash 2.4410$\pm$0.0140 &3036.74$\pm$110 & 0.165$\pm$ 0.033&\textendash 2.399$\pm$ 0.134 \\
GJ 905$^{b}$     &2930$\pm$60 & 0.189$\pm$0.0080 & \textemdash & 3102.27$\pm$117 &0.197$\pm$ 0.038 & \textemdash \\
  \hline
\end{tabular}
 \vspace{1ex} 
 
 \textbf{Notes.}
     \raggedright \small $^{a}$ Interferometrically measured parameters taken from Newton et al. (\citet{Newton2015}) Table 3. \\
     \raggedright $^{b}$ Parameters are from Mann et al. (\citet{Mann2015}). \\
     \raggedright $^{c}$ $T_{eff}$ is inferred using Equation (\ref{eqn22}); The fitting is shown in Figure \ref{fig:fig3}, top left. \\
     \raggedright $^{d}$ Radius is inferred using Equation (\ref{eqn23}); The fitting is shown in Figure \ref{fig:fig3}, top right. \\
     \raggedright $^{e}$ Luminosity is inferred using Equation (\ref{eqn24}); The fitting is shown in Figure \ref{fig:fig3}, bottom. \\
     
\end{table*}

\section{Result and Discussion :}
\subsection{H and K-Band EWs and $H_{2}O$ index} 

We have measured the equivalent widths (EWs) of atomic absorption lines due to Mg (1.57 $\mu$m), Al (1.67 $\mu$m), Mg (1.71 $\mu$m), Na (2.205 $\mu$m), Ca (2.263 $\mu$m) and the $H_{2}O$ spectral index in both the H- and K-bands. We used the H-band spectral features and water indices for the calibration of fundamental parameters like $T_{eff}$, stellar radius, luminosity, the H and K-band water indices to estimate the spectral type and absolute $K_{s}$ Magnitude ($M_{k_{s}}$). The K-band features and indices are used to determine the metallicities ([Fe/H] and [M/H]).  The equation to define the EW of an absorption line is :

\vspace*{-0.5cm} \begin{equation} \label{eq1}
\hspace*{2.5cm}  EW_{\lambda} = \int_{\lambda1}^{\lambda2} [1 - \frac{F(\lambda)}{F_{c}(\lambda)}] d\lambda
\end{equation}

Here, $F(\lambda)$ stands for the flux across the wavelength range of the line $(\lambda_{2} - \lambda_{1} )$, and $F_{c}(\lambda)$ is the estimated continuum flux on either side of the feature. All the EWs and water indices were calculated using slightly modified versions of the publicly available IDL-based \texttt{tellrv}\footnote{\url{https://github.com/ernewton/tellrv}} and \texttt{nirew}\footnote{\url{https://github.com/ernewton/nirew}} packages originally created by Newton et al. (\citet{Newton2014}, \citet{Newton2015}). As we know that the continuum of such cool type dwarf star's spectra are very much polluted by the molecular or broad absorption features, a pseudocontinuum is calculated using regions close to the feature of interest. To estimate the H and K-band EWs we choose EW Bandpasses and continuum windows very similar to those used by Rojas-Ayala et al. (\citet{RojasAyala2012}) and Newton at al. (\citet{Newton2015}) respectively with some minor modification where needed. The features with the continuum points used to calculate the equivalent widths are given in Table \ref{tab:table3_spectralfeature}. Based on the SNR of our spectra we have estimated the uncertainties on stellar parameters using a Monte Carlo simulation where multiple random realizations of gaussian noise are added to the spectrum and EWs of the features are recalculated. The final errors on the parameters were estimated by combining the randomly generated errors with the intrinsic scatters inherent in the calibration relations.\\
$~~~~~~~$ We have calculated $H_{2}O$-K indices for our observed M-dwarf spectra following Rojas-Ayala et al. (\citet{RojasAyala2012}):

\vspace*{-0.3cm} \begin{equation} \label{eq2}
\hspace*{0.5cm} H_2O-K2 = \frac{\langle{\mathcal{F}(2.070-2.090)}\rangle/\langle{\mathcal{F}(2.235-2.255)}\rangle}{\langle{\mathcal{F}(2.235-2.255)}\rangle/\langle{\mathcal{F}(2.360-2.380)}\rangle}
\end{equation}
where $\langle{\mathcal{F}(A-B)}\rangle$ denotes the median flux level in the wavelength range A to B, and a larger value of the index means a smaller amount of $H_{2}O$ opacity. We also used the H-band water index as defined by Terrien et al.\citet{Terrien2012} to estimate the calibration relations for $T_{eff}$ and radius:
\begin{equation} \label{eq3}
\hspace*{0.5cm} H_2O-H = \frac{\langle{\mathcal{F}(1.595-1.615)}\rangle/\langle{\mathcal{F}(1.680-1.700)}\rangle}{\langle{\mathcal{F}(1.680-1.700)}\rangle/\langle{\mathcal{F}(1.760-1.780)}\rangle}
\end{equation}
As for the EWs, $H_{2}O$ index uncertainties were estimated by adding synthetic random noise to our spectra consistent with their SNR, and taking the standard deviation of 100 synthesized $H_{2}O$ index values.

\begin{figure*}
\centering
\begin{tabular}{c}
\hspace*{-0.6cm} \includegraphics[width= 7.3 in, height = 1.8 in, clip]{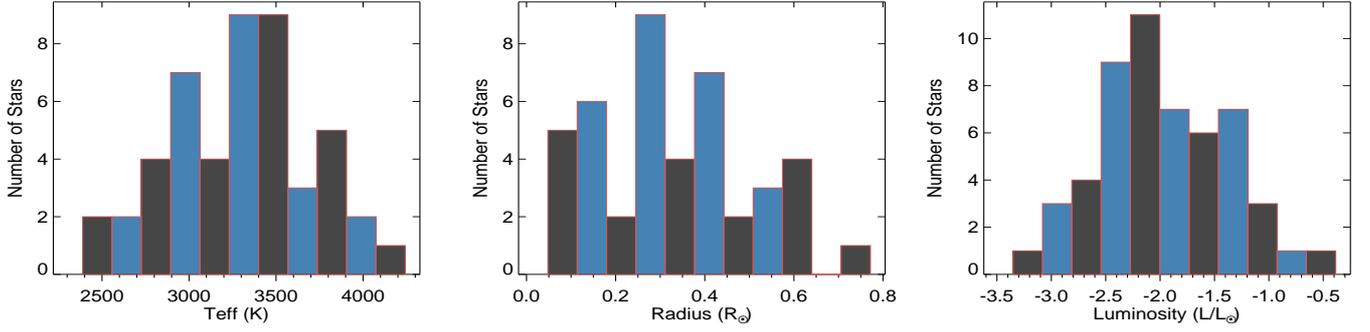}
\end{tabular}
\caption{The distribution of $T_{eff}$ (2390 < $T_{eff}$(K) < 4073 ), radius (0.05 < $R/R_{\sun}$ < 0.70) and luminosity (\textendash 3.35 < log $L/L_{\sun}$ < \textendash 0.66) for the M-dwarf stars of our observed sample are shown here.  }
\label{fig:fig4}
\end{figure*}

\subsection{Behavior of Spectral features: Comparison with model spectra}

We plot the H- and K-band EWs, EW ratio, and water indices against $T_{eff}$ for our calibrator sample (where interferometry has been used to determine parameters) in Figure \ref{fig:fig2}. Data points are color coded by their [Fe/H] values using the calibration relation in this work. We chose the new version of {\texttt{BT-Settl}}\footnote{\url{https://phoenix.ens-lyon.fr/Grids/BT-Settl/}} synthetic model spectra developed by Allard et al.\citet{Allard2014} and measured the exact spectral features to make a comparative analysis with our observed spectra. This new model predicts the NIR spectral distribution of M dwarfs with better accuracy as it's molecular line lists are updated and a new estimate of solar oxygen abundances are included as defined by Asplund et al. (\citet{Asplund2009}). We have degraded the resolution of the model spectrum to provide a better match to that of our observed spectra and measured the EWs and water indices numerically as described earlier. Figure \ref{fig:fig2} includes these over-plotted synthetic results for $T_{eff}$ = 2300K-4200K. For log g =5.0 we show the $T_{eff}$ tracks of sub solar (-0.5), solar (0.0) and super solar (+0.5) iron abundance [Fe/H] values, which spans the metallicity values of our calibrator samples. We chose two surface gravity values log g = 5.0 and log g = 4.5. The log g= 5.0 value is considered as the best match for empirical M dwarf gravity values (Fernandez et al. \citet{Fernandez2009}; Demory et al. \citet{Demory2009}).\\
$~~~~~~~~~$The Mg I (1.57$\mu$m) EWs predicted by the models show the best agreement with the observed spectra and it shows little metallicity dependence for $T_{eff}$ less than 3100K as noticed by Newton et al. (\citet{Newton2015}). The strength of Mg I (1.57 $\mu$m) increases with increasing $T_{eff}$ and metallicity. For solar metallicity, the strength of Mg I (1.57 $\mu$m) is decreasing with surface gravity (log g). The Al-a (1.67 $\mu$m) and Mg (1.71 $\mu$m) EWs behave in the same fashion as Mg (1.57 $\mu$m) though the strength for our observed calibrators is less than that of the model spectra. This could be due to the models, or relate to the binning procedure used to degrade the high resolution to that of the moderate resolution of HCT spectra. The nature of the Mg (1.57 $\mu$m) and Al-a (1.67 $\mu$m) equivalent width ratio is similar to what found in Newton et al. (\citet{Newton2015})'s work for $T_{eff}$ >3000K, though it shows strong metallicity dependence for $T_{eff}$  lower than 3000K and the values are higher for our calibration stars than that of the model spectra. \\
$~~~~~~~~~$The behavior of the Na I (2.205 $\mu$m) and Ca I (2.26 $\mu$m) feature and $H_{2}O$ indices for both bands are almost identical to what was found by Rojas-Ayala et al. (\citet{RojasAyala2012}). Observational evidence suggests that the strength of the Na I feature decreases as $T_{eff}$ increases until spectral type M6 ($T_{eff}$ $\sim$ 3000K; Cushing et al. \citet{Cushing2005}). This behaviour is seen for our solar ([Fe/H]=0) model spectra though it shows a jump near $\sim$ 3000K and then it's strength decreases with decreasing $T_{eff}$. For the sub-solar model (-0.5dex) the strength is maximum at 3000K and then it tends to decrease for both the hotter and cooler ends. The super-solar model ([Fe/H]=0.5) shows strikingly different behaviour where the strength weaken with decreasing $T_{eff}$. The abnormality in the nature of the model Na I EWs could be due to the formation of molecular species at low $T_{eff}$ (Rojas-Ayala et al. \citet{RojasAyala2012}). The strength of the Ca I feature for the synthetic spectra increases both with $T_{eff}$ and metallicity. The H and K-band $H_{2}O$ indices increase monotonically with increasing $T_{eff}$ (larger value of index means less water absorption); this change is almost independent of metallicity effects. 

\begin{figure*}
\centering
\begin{tabular}{cc}
\hspace*{0.5 cm} \includegraphics[width= 5.8 in, height = 2.2 in, clip]{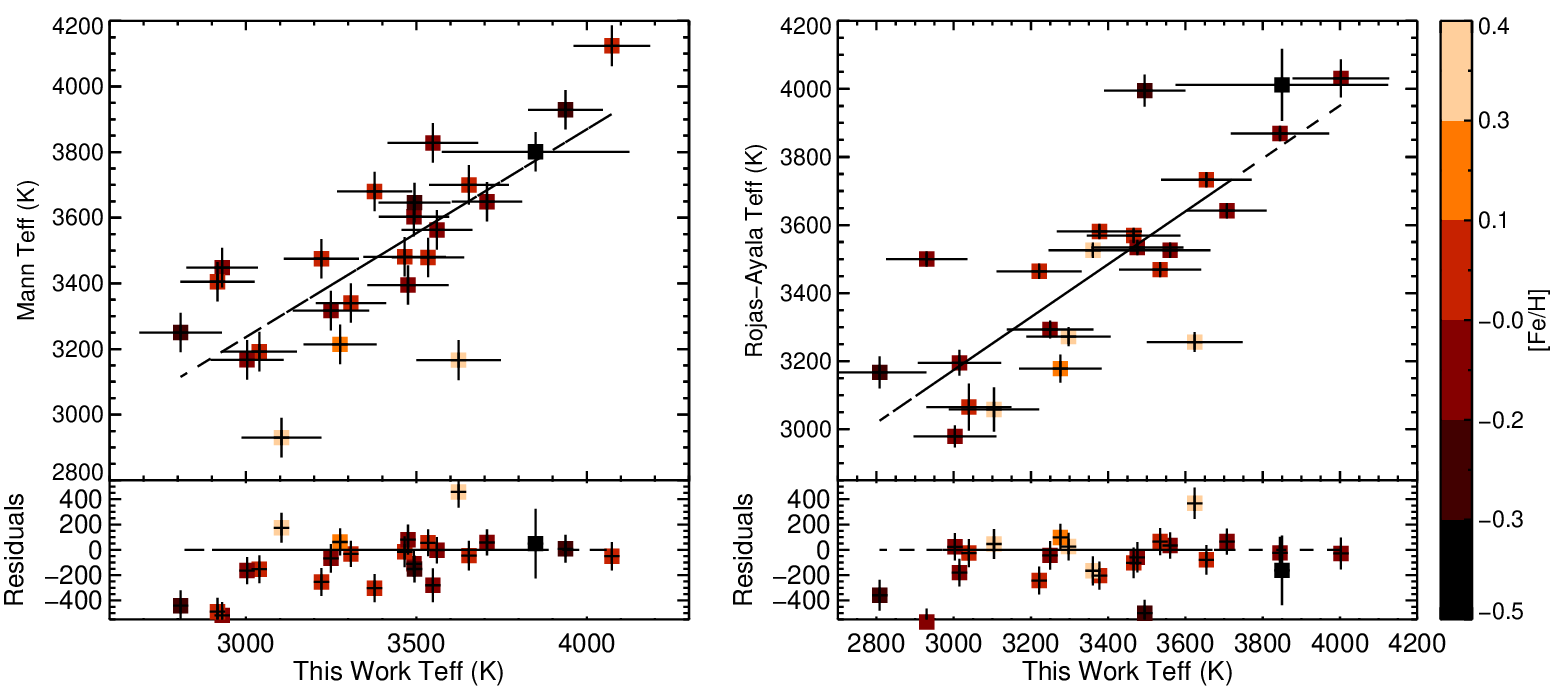}\\[2\tabcolsep]
\hspace*{0.5 cm} \includegraphics[width= 5.8 in, height = 2.2 in, clip]{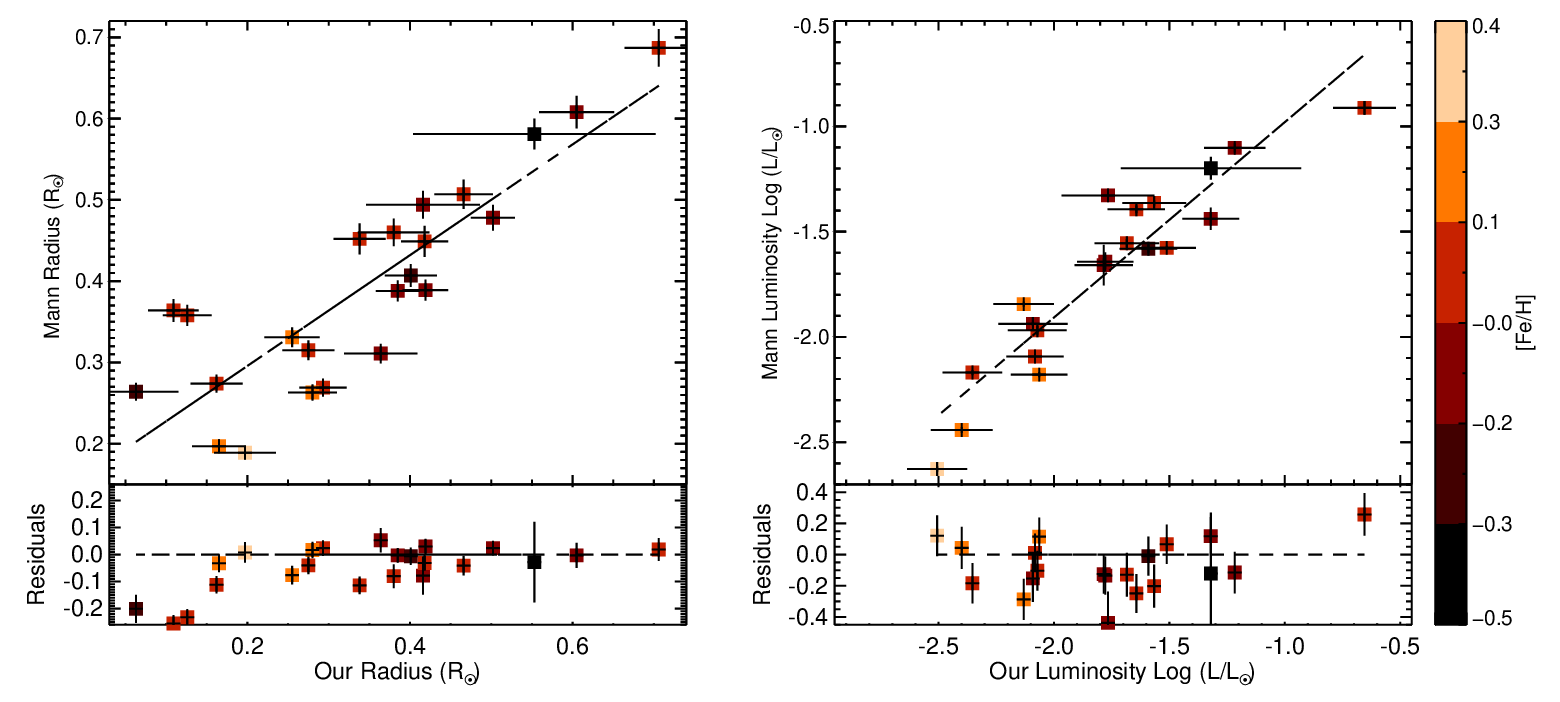}
\end{tabular}
\caption{Comparisons of the spectral parameters of M dwarfs derived in this work (shown in horizontal axes) with different methods used in other work (shown in vertical axes) with residuals in the bottom panels are plotted. The top left and right plots represent the comparison of our estimated $T_{eff}$ with the values derived using relations from Mann et al. (\citet{Mann2015}) and Rojas-Ayala et al. (\citet{RojasAyala2012}). The bottom left and right are the comparisons of radius and luminosity with work done by Mann et al. (\citet{Mann2015}). All the data points are color coded by the [Fe/H] values measured in this work.}
\label{fig:fig5}
\end{figure*}

\begin{figure*}
\centering
\begin{tabular}{c}
\hspace*{-0.5cm} \includegraphics[width= 7.2 in, height = 2.3 in, clip]{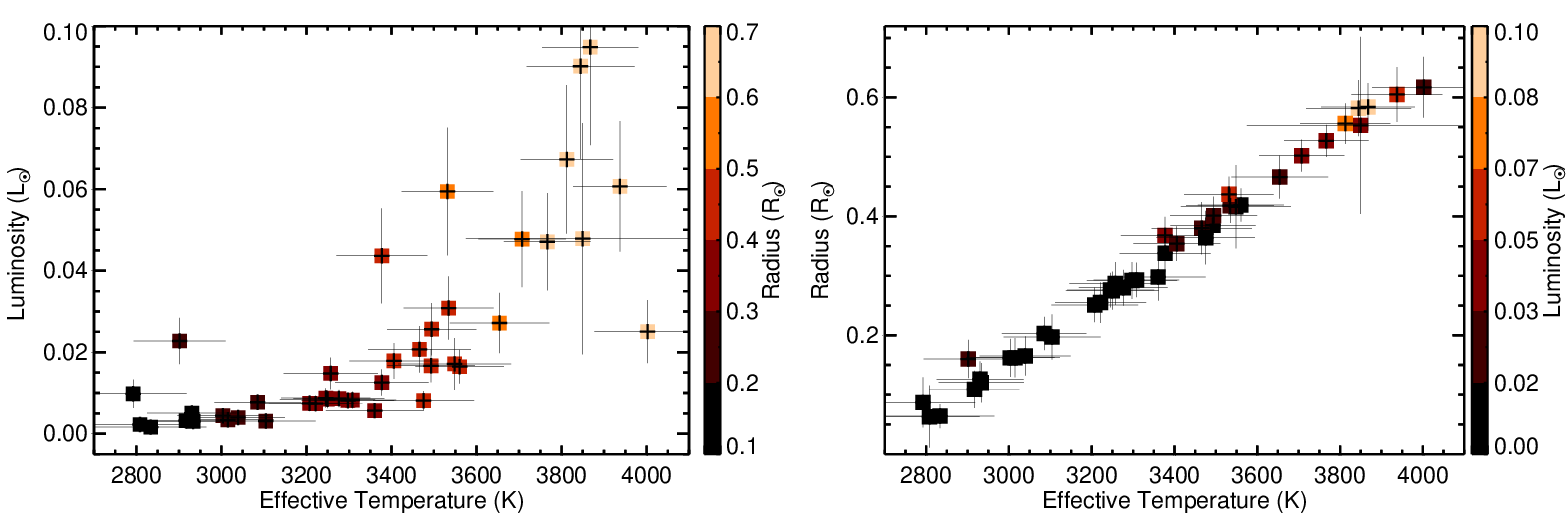}
\end{tabular}
\caption{\textbf{Left:} Stellar luminosity vs. $T_{eff}$ estimated using EW dependent calibration relation. \textbf{Right:} Stellar radius vs. $T_{eff}$. The points are shaded with colors according to the derived stellar radii/luminosities.}
\label{fig:fig6}
\end{figure*}

\subsection{Empirically Calibrating $T_{eff}$, Radius and Luminosity} 
Visual inspection of the H-band spectra of M-dwarfs supports the fact to a large extent that there exists a clear indication of the dependence of parameters like $T_{eff}$, stellar radius, luminosity with spectral features. Using stars with interferometrically measured parameters as calibrators, Newton et al. (\citet{Newton2015}) established linear calibration relations for M dwarf samples as a function of EWs, EW ratios of the relatively most robust spectral features. The H-band spectra obtained with the TIRSPEC instrument of the 2-m HCT did not cover the Mg I (1.50 $\mu$m) feature due to instrumental cut-off, so it is absent in the calibration relations for $T_{eff}$ and luminosity. Newton et al. (\citet{Newton2015}) conducted a principal component analysis (PCA) test to identify the EWs of those features which have the strongest correlations with $T_{eff}$, radius, and luminosity. The authors explored simple parameterizations of one or multiple EWs or EW ratios and found the best fits which have the lowest Bayesian Information Criterion (BIC). Following those established correlations, the main target of our work was to also consider the H-band water index, and find a new set of calibration rules. We checked with different single-line and multi-line functional forms of selected H-band EWs, EW ratios, and water index. We performed a multivariate linear regression to determine the best-fitting parameters for different possible combinations of features or indices. We used the adjusted square of the multiple correlation coefficient ($R^{2}_{ap}$) (Rojas-Ayala et al. \citet{RojasAyala2012}), the RMSE and the mean absolute deviation (MAD) values to qualitatively compare the goodness of each fit. The RMSE evaluates the accuracy and predictive power of the response of the regression model, and the $R^{2}_{ap}$ depicts the proportion of variability in a data set that is accounted for by a regression model. For a given set of calibration data, lower RMSE value and $R^{2}_{ap}$ value closer to 1 indicate better fit. Now the study of the behavior of Mg (1.57 $\mu$m) EWs shows that the feature correlates strongly with $T_{eff}$, and it closely follows the temperature tracks of Mg (1.50 $\mu$m) feature. We found that the choice of the EW of Mg (1.57 $\mu$m), in the place of the EW ratio of Mg (1.50 $\mu$m) and Al-b (1.67 $\mu$m) in the calibration relation of $T_{eff}$ significantly improves the fit with the highest $R^{2}_{ap}$ value. Similarly, in the luminosity relation, the inclusion of Mg (1.57 $\mu$m) gave a better $R^{2}_{ap}$ value than what we got for the other combinations of EWs or EW ratios. Our best fitting calibration relations for all three stellar parameters are given as:

\vspace*{-0.2cm} \begin{equation}
\begin{split}
\label{eqn22}
\hspace*{0.2cm} T_{eff}/K &=\ +592.94 - 364.34\times Al-a (1.67 \mu m)+224.07  \\
\hspace*{0.2cm}             &\ \    \times Mg (1.57 \mu m) + 2936.78\times H_{2}O{\text -}H \\
\end{split}
\end{equation}
\hspace*{0.8cm} RMSE ($T_{eff}$/K) = 102 \\
\hspace*{1.7cm} MAD / K = 64

\vspace*{-0.2cm} \begin{equation}
\begin{split}
\label{eqn23}
\hspace*{0.3cm} R/R_\odot\ &=\ -0.854 + 0.119\times Mg (1.57 \mu m)-0.150\times \\
\hspace*{0.3cm}             &\ \ Al-a (1.67 \mu m) - 0.004 \times Mg (1.57 \mu m)/ \\
\hspace*{0.3cm}             &\ \  Al-a (1.67 \mu m) + 1.203\times H_{2}O{\text -}H  \\
\end{split}
\end{equation}
\hspace*{0.8cm} RMSE ($R/R_{\sun}$) = 0.027 \\
\hspace*{1.14cm} MAD / $R_{\sun}$ = 0.017

\begin{equation}
\begin{split}
\label{eqn24}
\hspace*{0.3cm} log L/L_\odot\ &=\ -2.888 + 0.316\times Mg (1.71 \mu m)-0.101\times \\
\hspace*{0.3cm}             &\ \ [Mg (1.71 \mu m)]^{2} + 0.379 \times Mg (1.57 \mu m) \\
\end{split}
\end{equation}
\hspace*{0.8cm} RMSE ($log L/L_{\sun}$) = 0.12 \\
\hspace*{1.75cm} MAD / dex = 0.09  \\

\hspace*{-0.35cm} While performing multivariate linear regression on the chosen features to find the best-fitting relationships with the parameters, the inclusion of the H-band water index in the relations of $T_{eff}$ and radius gave us a better result. Numerous earlier observations predict that the overall shape of M dwarf spectra both in H- and K-band changes due to $H_2$O opacity. The water absorption is sensitive with spectral type (Kleinmann et al. \citet{Kleinmann1986}; McLean et al. \citet{McLean2003}) and $T_{eff}$ (Merrill et al. \citet{Merrill1979}; Jones et al. \citet{Jones1994}; Ali et al. \citet{Ali1995}; Rojas-Ayala et al. \citet{RojasAyala2012}). As it influences the strength of the spectral features greatly, we have introduced $H_2$O-H index defined by Terrien et al. \citet{Terrien2012} in our calibration relations for $T_{eff}$ and stellar radius. Now, in terms of the residuals, the calibration relations that are given by Newton et al. (\citet{Newton2015}) in previous work for $T_{eff}$ and luminosity might appear to be a little better than this work. The reason could be as the SpeX instrument on IRTF provides superior quality of spectra with better resolution (R = 2000) and higher signal to noise ratio (SNR $\geq$ 100) than that of the TIRSPEC spectra (R = 1200). However, our estimated parameters cover a wider range of properties, and the statistical analysis of the fits supports that the inclusion $H_2$O-H indices improve the fits to a significant extent. In support of our claim we can say that without the $H_2$O-H index in the radius relation (which is very similar to the Newton relation), the residual and $R^{2}_{ap}$ values are RMSE(Radius) = 0.037$R_{\sun}$, $R^{2}_{ap}$(Radius) = 0.94; while with the introduction of the $H_2$O-H index, the internal quality of the fit has improved with much lower RMSE(Radius) = 0.027$R_{\sun}$ and higher $R^{2}_{ap}$(Radius) = 0.97 values. The addition of the water index in the calibration relation of $T_{eff}$ also provide better estimation of result with the RMSE($T_{eff}$) value reducing from 216K to 102K and the $R^{2}_{ap}$($T_{eff}$) value improving from 0.53 to 0.89. Furthermore, for a given relatively low-resolution spectra compared to the previous work (Newton et al. \citet{Newton2015}), our calibrations provide reliable results within an acceptable limit of accuracy estimation. We have not included the water index for luminosity calibration ($R^{2}_{ap}$(log L/$L_{\sun}$) = 0.93) as we did not get significant improvement in our result.

$~~$ In Figure \ref{fig:fig3} we show our best fits for the calibrator stars with residuals of the respective fits and in Table \ref{tab:table4_teff_rad_lum_cal} we present the inferred stellar parameters with respective errors that we obtained using those calibration relations. We estimated the uncertainties in our stellar parameters by adding the random gaussian error propagated from the features and indices added with the standard deviation of the residuals of our best fits. We show the distribution of spectral parameters of our observed M-dwarf stars in Figure \ref{fig:fig4} and our sample lies in the parameter space of 2390K $\textless T_{eff} \textless$ 4073K, 0.05$R_{\sun} \textless R \textless 0.7R_{\sun}$ and - 3.35 $\textless log (L/L_{\sun}) \textless $ - 0.66.

\begin{figure*}
\centering
\begin{tabular}{c}
\hspace*{-0.95cm} \includegraphics[width= 7.7 in, height = 2.1 in, clip]{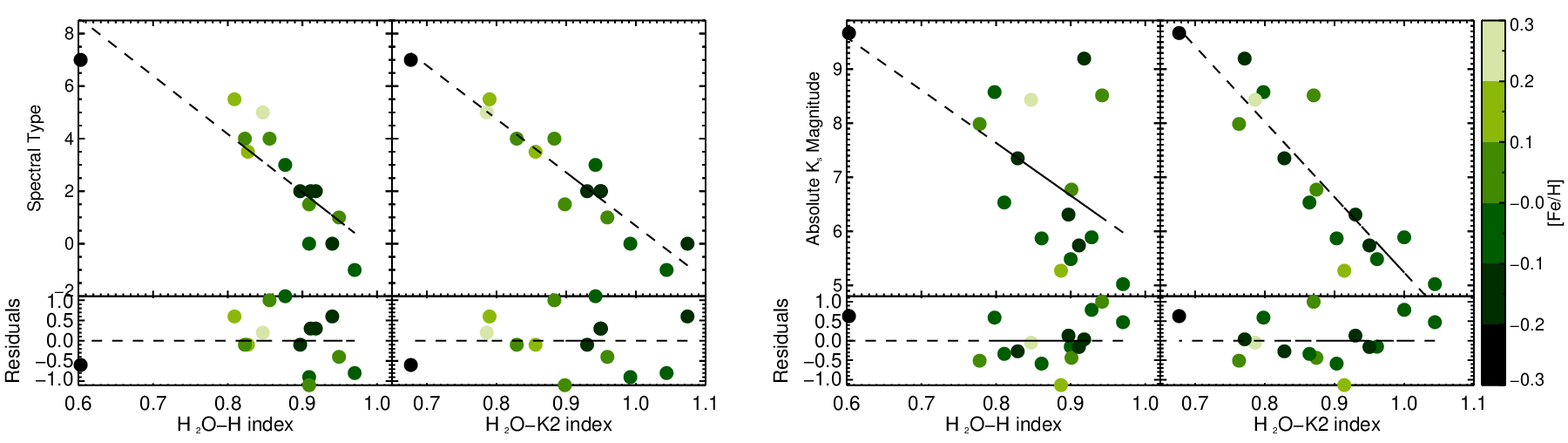}
\end{tabular}
\caption{\textbf{Left:} M-dwarf spectral types vs. H and K-band $H_{2}O$ index for our calibration samples to measure the spectral types. \textbf{Right:} Absolute $K_{s}$ magnitude ($M_{k_{s}}$) vs H and K-band $H_{2}O$ indices for our calibrator stars. For both of the plots, the dashed black lines represent the linear relationships with the water indices. The data points are color-coded according to their metallicity [Fe/H] values.}
\label{fig:fig7}
\end{figure*}

\begin{figure}
\centering
\begin{tabular}{c}
\hspace*{-0.4 cm} \includegraphics[width= 3.5 in, height = 2.4 in, clip]{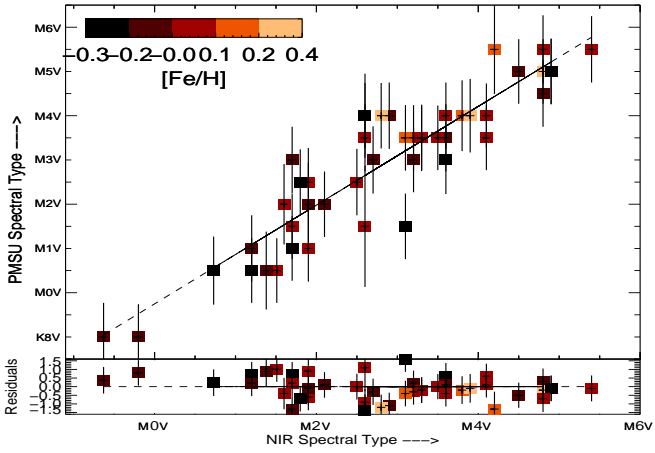}
\end{tabular}
\caption{Comparison of NIR Spectral type estimated from the linear relation using H- and K-band $H_{2}O$ indices with PMSU Spectral types (Reid et al. \citet{Reid1995}; Hawley et al. \citet{Hawley1996}) for our overlapping M dwarfs. The color of the data points represents the metallicity [Fe/H] of the stars.}
\label{fig:fig8}
\end{figure}

\begin{table}
 \centering
 \caption{M Dwarf Spectral Type calibration samples}
 \label{tab:table5_sptype_cal}
 \begin{tabular}{lccccc}
  \hline
Star & $H_{2}O$-H & $H_{2}O$-K & Sp. Type & Sp.Type & Ref. \\
   &  &  & (Inferred) & (Standards) & \\
 \hline
HD232979  & 0.940 & 1.074 & K7 & M0 & c \\
HD79210   & 0.970 & 1.044 & M0.0 & K7 & a,b \\
Gl338B    & 0.909 & 0.992 & M1 & M0 & a,b \\
HD233153  & 0.949 & 0.959 & M1.4 & M1 & b \\
HD36395   & 0.909 & 0.898 & M2.6 & M1.5 & b \\
HD95735   & 0.897 & 0.930 & M2.1 & M2 & b \\
VGXAnd    & 0.918 & 0.949 & M1.7 & M2 & b \\
GL436     & 0.877 & 0.942 & M1.9 & M3 & a,b \\
HD119850  & 0.911 & 0.950 & M1.7 & M2 & c \\
GJ 3378   & 0.856 & 0.883 & M3.1 & M4 & c \\
GJ 3379     & 0.827 & 0.856 & M3.6 & M3.5 & c \\
Gl447     & 0.823 & 0.829 & M4.1 & M4 & a,c \\
Gl905     & 0.847 & 0.786 & M4.8 & M5 & b \\
Gl473   & 0.809 & 0.790 & M4.9 & M5.5 & a \\
GAT1370   & 0.603 & 0.677 & M7.6 & M7 & b \\
  
  \hline
\end{tabular}
 \vspace{1ex} 

     \raggedright \small $^{a}$ KHM Spectral Standards Kirkpatrick et al. (\citet{Kirkpatrick1991}, \citet{Kirkpatrick1995}, \citet{Kirkpatrick1999}).\\
     \raggedright $^{b}$ CARMENES input catalogue of M dwarfs (Alonso-Floriano et al. \citet{AlonsoFloriano2015}). \\
      \raggedright $^{c}$ CARMENES M dwarfs velocities (
A. Reiners et al., \citet{Reiners2018})
\end{table}

$~~$ In Figure \ref{fig:fig5} we show the comparison plots of the fundamental parameters determined in other work (vertical axes) following different methods with our work (horizontal axes) with residuals for the overlapping stars. In the top panel, we show the estimated $T_{eff}$ obtained using our relations against those from the Mann et al. (\citet{Mann2015}) and Rojas-Ayala et al. (\citet{RojasAyala2012}). While Mann et al. (\citet{Mann2015}) calculated the $T_{eff}$ by comparing the star's optical spectra with BT-Settl (PHOENIX) models, Rojas-Ayala et al. (\citet{RojasAyala2012}) inferred the $T_{eff}$ from the $H_{2}O$-K2 index using solar ([M/H]=0.0) BT-Settl models of K-band spectra. The differences in the median and standard deviation values of $T_{eff}$ are 110K and 164K respectively, between the present work and Mann et al. (\citet{Mann2015}), while those are 79K and 158K respectively, for Rojas-Ayala et al.  (\citet{RojasAyala2012}). In the bottom panel, we compare stellar radius (inferred using $T_{eff}$-metallicity-radius relation) and luminosity (inferred using $T_{eff}$-luminosity relation) obtained from Mann et al. (\citet{Mann2015}) with our findings for the common stars. The radii differences have a median value of $0.03R_{\odot}$ and a standard deviation of $0.07R_{\odot}$ and for the luminosity differences these values are respectively 0.13$L_{\odot}$ and 0.10$L_{\odot}$ in log scale. Though the plots display some scatter, it also shows good agreement with our estimated results. In the left panel of Figure \ref{fig:fig6}, we show the estimated luminosity as a function of $T_{eff}$ for our observed M dwarf stars with the data points are color-coded with their radius. In the right panel of Figure \ref{fig:fig6}, we show the stellar radius as a function of $T_{eff}$ for our work with the data points are shaded with colors according to their luminosity values. These two plots confirm one to one correlation of stellar luminosity and radius with $T_{eff}$. 

\subsection{Calibrating the $H_{2}O$ indices to estimate Sp. Type and (M$_{k_{s}}$)}

\subsubsection{\textbf {NIR Spectral Type:}} 

\begin{figure*}
\centering
\begin{tabular}{ccc}
\hspace*{-0.3cm} \includegraphics[width= 3.1 in, height = 2.3 in, clip]{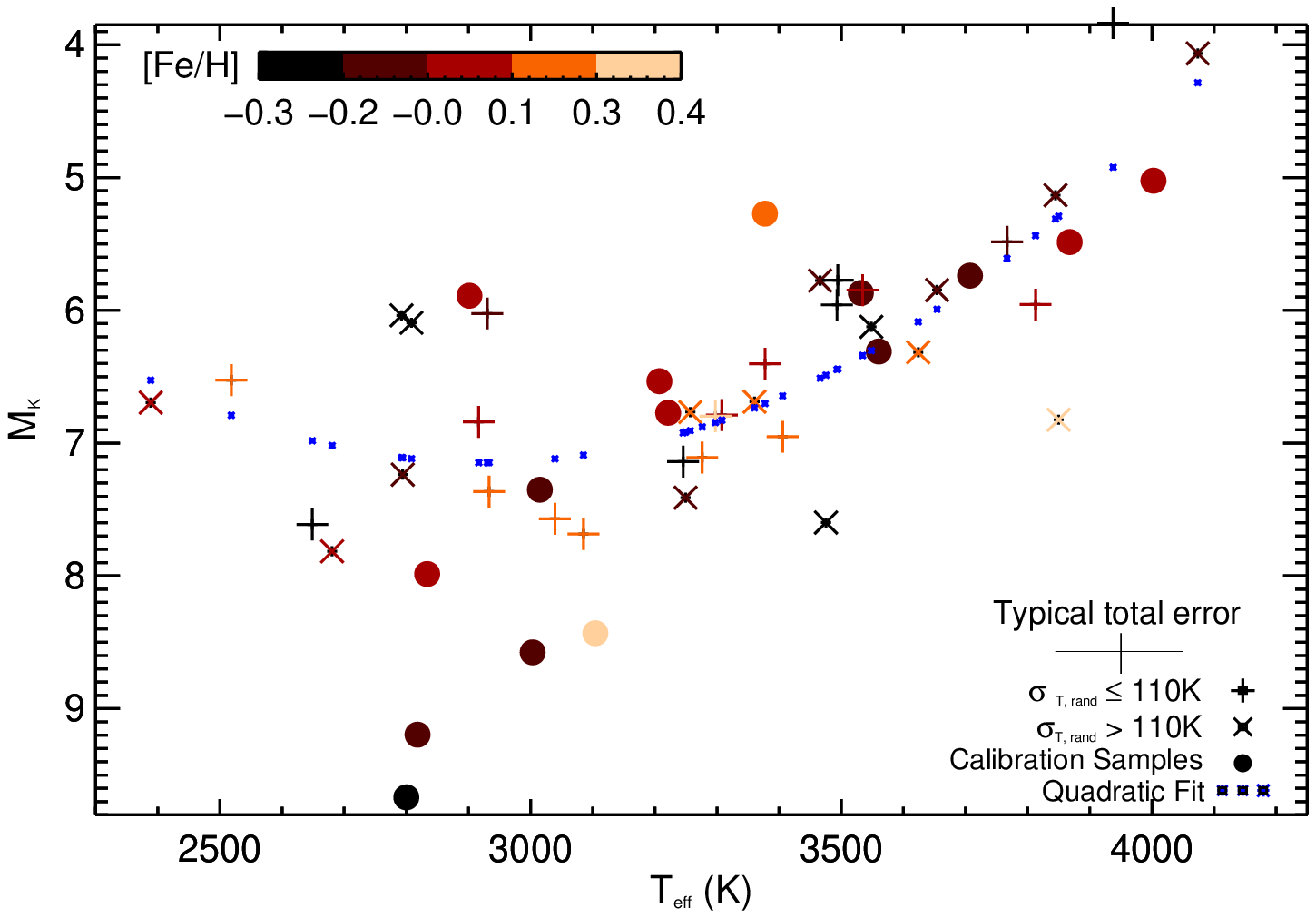}  \hspace*{0.8cm}
\includegraphics[width= 3.1 in, height = 2.3 in, clip]{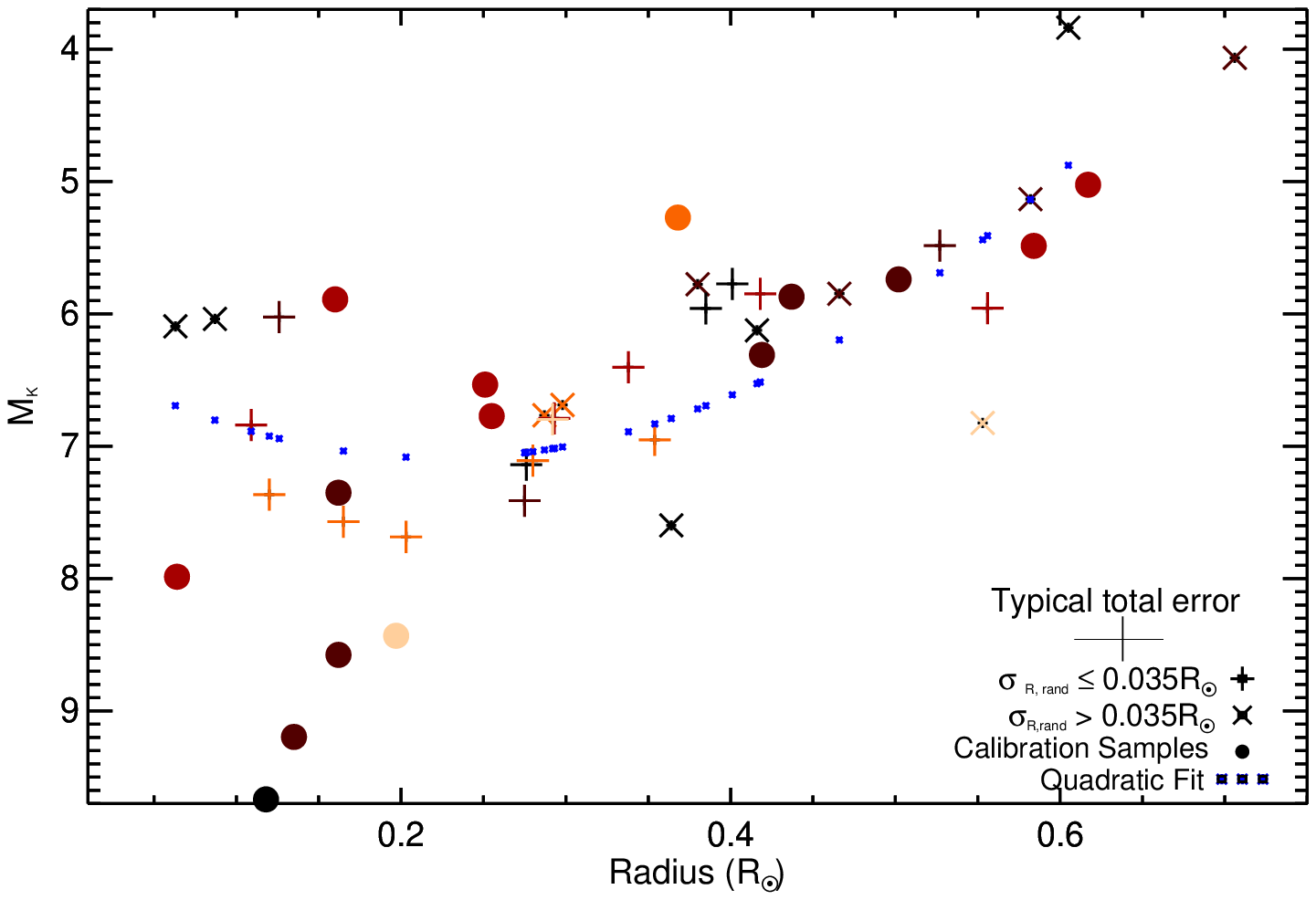}\\[2\tabcolsep]
\includegraphics[width= 3.1 in, height = 2.3 in, clip]{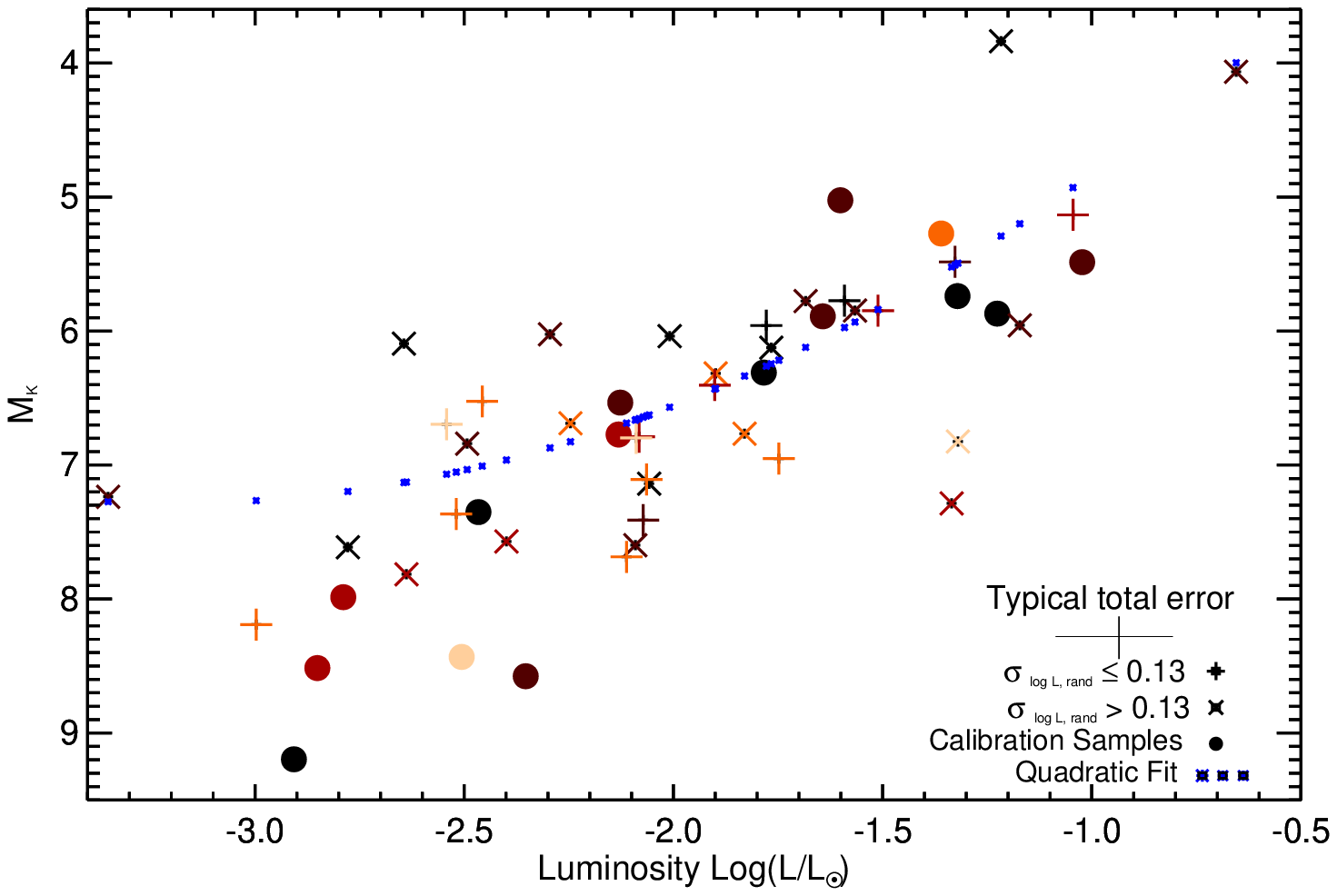}
\end{tabular}
\caption{Plot of absolute K$_{s}$ magnitude vs. inferred $T_{eff}$ , radius, or Log($L_{bol}$) for our 2-m HCT M-dwarfs samples. We use plus symbol for smaller errors, and crosses for larger errors for respective parameters. The filled circles represent the stars we use as calibrators. We also overplot the quadratic fits of M$_{k}$ with the parameters showing by blue dots. All stars are color-coded by their metallicity [Fe/H] values.}
\label{fig:fig9}
\end{figure*}

\begin{table}
 \centering
 \caption{List of objects used as Calibrators to estimate M$_{k}$.}
 \label{tab:table6_mag_cal}
 \begin{tabular}{lccc}
  \hline
  Star & K$_{s}$ $^{a}$ & Parallax $^{b}$ & M$_{k}$ \\
   & (mag) & (mas) & (mag)\\
  \hline
  HD79210 	& 4.032	  	& 157.88	& 5.024	\\
  GJ3395 		& 7.063		& 48.38	& 5.486	\\
  GJ 514 	& 5.300		& 131.24	& 5.890	\\
  GJ 625 		& 5.833		& 154.08	& 6.772	\\     
  HD95735 	& 3.340	   	& 392.64	& 6.310	\\
  GJ 494 	& 5.578		& 86.86	& 5.272	\\
  HD119850 	& 4.415		& 183.99	& 5.739	\\       
  GJ48 		& 5.449		& 121.42	& 5.870	\\       	
  GJ 3522 		& 5.688		& 147.66	& 6.534	\\       
  GJ445 		& 5.954		& 190.26	& 7.351	\\      
  GJ1156 		& 7.570		& 154.51	& 8.515	\\       	
  Gl166C 		& 5.962		& 333.33	& 8.576	\\
  Gl905 		& 5.929		& 316.70		& 8.432	\\
  Gl406 		& 6.084		& 419.10		& 9.196	\\
  GJ 1083 	& 8.039		& 97.60		& 7.986	\\
  GAT1370 	& 7.585		& 261.01	& 9.668	\\
  
  \hline
 \end{tabular}
 \vspace{1ex} 

     \raggedright \small $^{a}$ Apparent K$_{s}$ magnitude are from 2MASS (Cutri et al. \citet{Cutri2003}).\\
     \raggedright $^{b}$ References for Parallaxes are Gaia Collaboration et al. (\citet{Gaia2016}) and (\citet{Gaia2018}), van Leeuwen (\citet{vanLeeuwen2007}), Henry et al. (\citet{Henry2006}), Gatewood at al.\citet{Gatewood2008}, Dittmann et al.\citet{Dittmann2014}
\end{table}

Kleinmann and Hall (\citet{Kleinmann1986}) stated first that the $H_{2}O$ absorption in the spectra of dwarf stars causes the depression at K-band and the spectral type decreases as the water absorption increases. After that McLean et al. (\citet{McLean2003}) gave a linear relationship between NIR K-band water index and spectral type for M, L, and T type dwarfs. Most recently, Rojas-Ayala et al. (\citet{RojasAyala2012}) used  $H_{2}O$-K2 index as spectral type indicator and established a linear relationship using the KHM (established by Kirkpatrick et al. \citet{Kirkpatrick1991}, hereafter KHM) spectral standards (Henry at al. \citet{Henry2006}). In this work, we took both H and K-band $H_{2}O$ index and obtained a calibration relation with spectral type:
\begin{equation}
\begin{split}
\label{eqn34}
\hspace*{0.2cm} {Sp}_{NIR}  &= \ 21.68 - 3.45\times (H_{2}O{\text -}H) - 17.76\times (H_{2}O{\text -}K) \\
\end{split}
\end{equation}
\hspace*{0.5cm} RMSE ($Sp_{NIR}$) = 0.73 \\
\hspace*{0.65cm} MAD ($Sp_{NIR}$) = 0.55

We used 15 stars as calibrators (Table. \ref{tab:table5_sptype_cal} ) to identify the best-fit relationship. The uncertainties in Sp. types for our stars are calculated from the random error in the $H_{2}O$ indices by an MCMC (Markov chain Monte Carlo) simulation using \texttt{idl\textunderscore emcee}\footnote{\url{https://github.com/mcfit/idl_emcee}}. We used 500 walkers and 100 iterations where the walkers are initialized as a gaussian over the specified range between the min and max values of each free parameter within the 3$\sigma$ confidence level. We determined the final error by adding the randomly generated Gaussian errors with the intrinsic scatter of the best-fit relation. As all of our calibrators did not belong to the KHM  system, we used the CARMENES catalogue of M dwarfs and took the measured spectral types from Low-resolution optical spectroscopy (F. J. Alonso-Floriano et al. \citet{AlonsoFloriano2015}) and High-resolution optical and NIR spectroscopy (A. Reiners et al. \citet{Reiners2018}) to establish the calibration relation. Our Calibration relation is valid roughly for spectral types ranging from K7V-M7V, and the RMSE of our relationship is around 0.73 which indicates that we can tell each star's sp. types with an accuracy of the order of 0.7 subtypes and the $R^{2}_{ap}$(sp.type) value is 0.90 which predicts the quality of the fits. In Figure \ref{fig:fig8}, we show a comparison plot of our calibrated spectral type with PMSU (Palomar/MSU nearby star spectroscopic survey; Hawley et al. \citet{Hawley1996}) spectral types. The spectral type differences have a median value of 0.4 subtypes and a standard deviation of 0.44 subtype for the PMSU vs. our estimation. For spectral type notation we expressed M5V numerically as $Sp_{NIR}$ = 5 and for K7V we took as $Sp_{NIR}$ = -1.0.

\begin{figure*}
\centering
\begin{tabular}{cc}
\hspace*{-0.95cm} \includegraphics[width= 2.7 in, height = 3.7 in, angle=-90, clip]{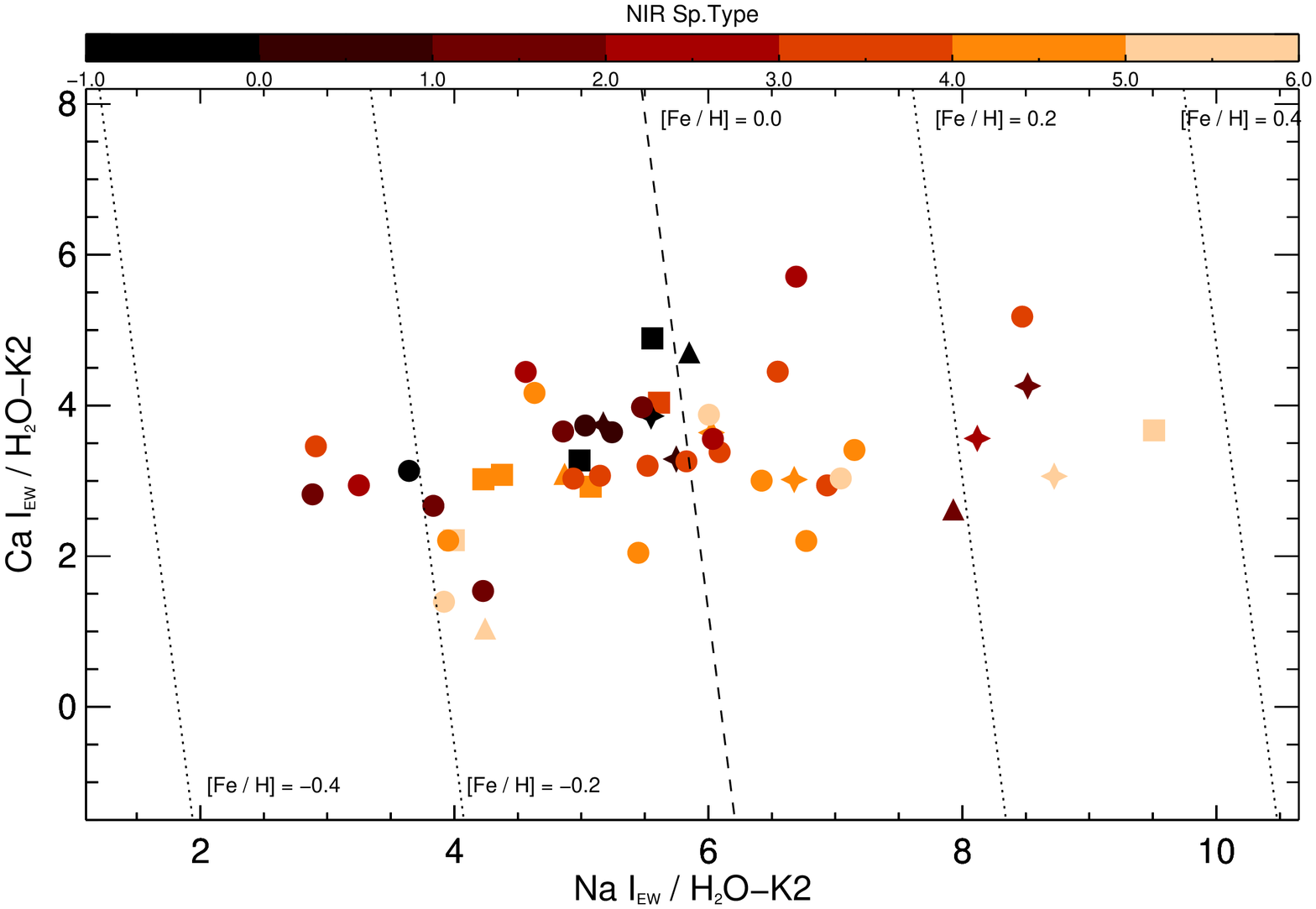} \hspace*{0.2cm}
\includegraphics[width= 2.8 in, height = 3.5 in, angle=-90, clip]{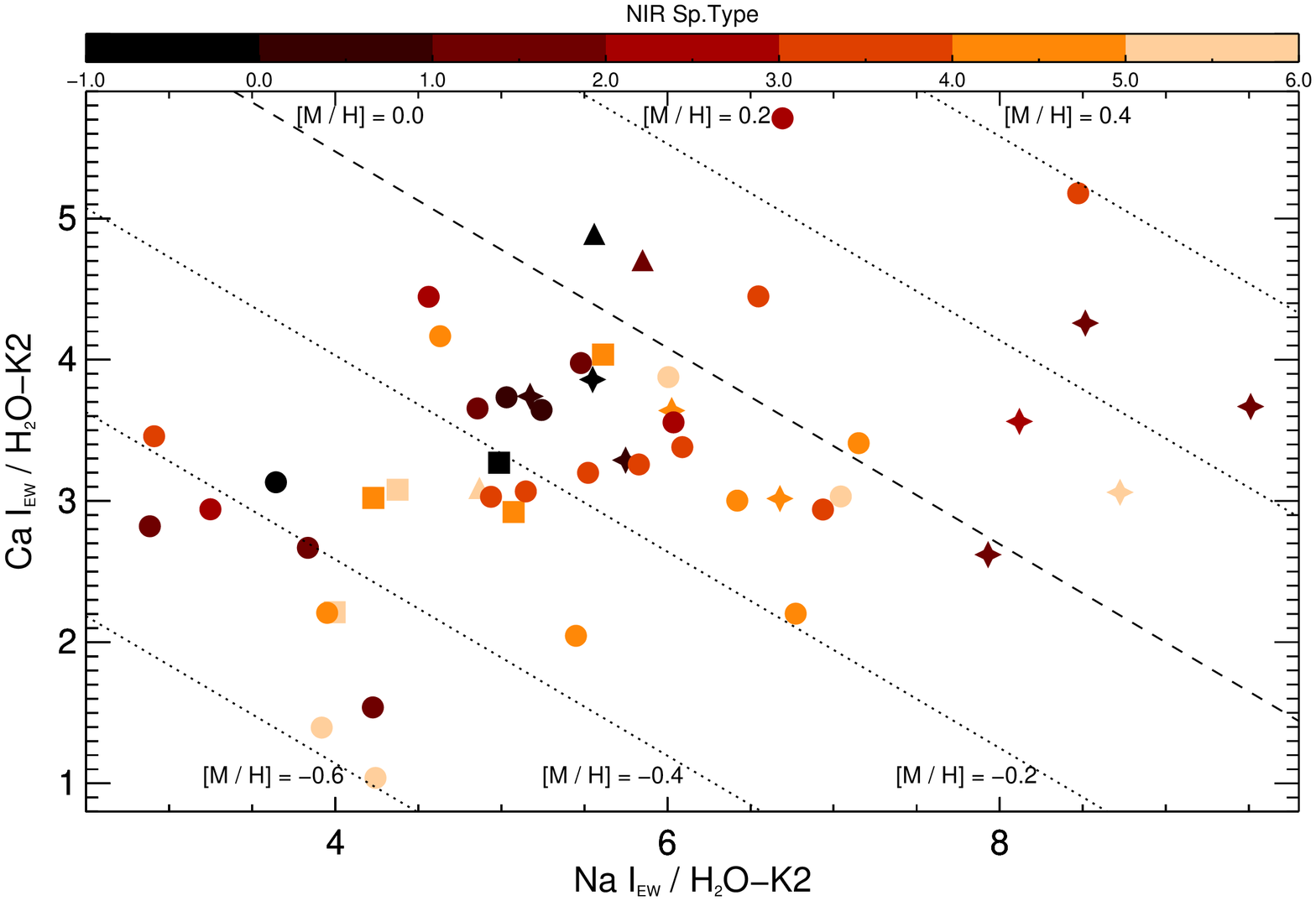}
\end{tabular}
\caption{\textbf{Left}: The plot of CaI EW vs. NaI EW, both weighted by the $H_{2}O$-K2 index, for the observed M dwarfs spectra. The filled triangles and squares
are the calibration samples with [Fe/H] $\geq$ 0.0 and [Fe/H] < 0.0 respectively. The dashed and dotted lines are the isometallicity contours (calculated from Equation \ref{eqn56}) for different values of [Fe/H]; +0.4, +0.2, +0.0, \textendash 0.2 and \textendash 0.4 dex from the top right to bottom left corner, respectively. \textbf{Right}: This plot is same as the left one, but here the filled triangles and squares are the calibration samples with [M/H] $\geq$ 0.0 and [M/H] < 0.0 respectively. The dashed and dotted lines are the isometallicity contours (calculated from Equation \ref{eqn58}) for different values of [M/H]; +0.4, +0.2, +0.0, \textendash 0.2, \textendash 0.4 and  \textendash 0.6 dex from the top right to bottom left corner, respectively.
For both of the [Fe/H] and [M/H] plots, the star-shaped are potential planet hosts, and the round dots are the rest of the stars. Majority of the planet host stars are super solar-type with metallicity > 0.0 dex (some are within \textendash 0.05 to 0.0 dex). Both of the plots are color-coded by spectral types.}
\label{fig:fig10}
\end{figure*}

\subsubsection{\textbf {Absolute $K_{s}$ Magnitude and Spectroscopic Distances}}

Newton at al. (\citet{Newton2014}) presented a calibration relation for estimating absolute K$_{s}$ magnitude using NIR spectral type and the $H_{2}O$-K2 index. Now for our observed dwarf stars, we obtained an improved calibration relation by adapting both the $H_{2}O$ indices to get a linear fit with absolute $K_{s}$ magnitude. First we calculated M$_{k}$ for our calibrators (Table. \ref{tab:table6_mag_cal}) taking apparent K$_{s}$ magnitudes from 2MASS and respective parallaxes. Our best fits are valid for the index ranges  0.60 \textless$H_{2}O$-H \textless 0.97 and 0.67 \textless $H_{2}O$-K2 \textless 1.08 :

\begin{equation}
\hspace*{0.9cm} M_{k} = 17.74 + 5.76(H_{2}O{\text -}H) - 17.99(H_{2}O{\text -}K2)
\label{eqn111} 
\end{equation}
\hspace*{0.8cm} RMSE ($M_{k}$) = 0.62 \\
\hspace*{0.93cm} MAD ($M_{k}$) = 0.44

\hspace*{-0.7cm} The RMSE of our regression model fit is 0.62 and the $R^{2}_{ap}$(M$_{k}$) = 0.83. We also estimated the spectroscopic distance using M$_{k}$ values and apparent magnitudes (Cutri et al. \citet{Cutri2003}), ignoring the extinction effects. The uncertainties in absolute K$_{s}$ magnitude and distance are estimated from the random errors in the $H_{2}O$ indices following the same method as we applied for sp. type. The estimated M$_{k}$ values and spectroscopic distances are given in Table \ref{tab:table9_finalresult}.

\subsubsection{\textbf {Relation between M$_{k}$ and inferred parameters}}
We show the plot of absolute K$_{s}$ magnitude (M$_{k}$) vs. inferred $T_{eff}$, stellar radius and luminosity (log $L_{bol}$) respectively in Figure \ref{fig:fig9}. All the parameters are estimated using NIR spectroscopic features and indices. For very low mass stars, Delfosse et al. (\citet{Delfosse2000}) established visual and NIR empirical mass-luminosity relations. As most of the stellar properties have a very close mass dependency, it is very obvious that there exist strong relationships between M$_{k}$ and other fundamental parameters. So we can consider M$_{k}$ as an independent indicator of stellar parameters (Newton at al. \citet{Newton2015}) and we also show a quadratic fit of M$_{k}$ as a function of $T_{eff}$, stellar radius and luminosity ($L_{bol}$) respectively :

\begin{eqnarray}
\label{eqn36}
M_{K}&=& -11.19 + 0.013\times T_{eff} - 2.140\times 10^{-6}\times T_{eff}^{2}\\
\label{eqn46}
&=& 6.22 + 6.03\times R_\star\ - 13.46\times R^{2}_\star\ \\
\label{eqn56}
&=& 2.46 - 3.09\times log (L/L_\odot) - 0.48\times log (L/L_\odot)^{2}       
\end{eqnarray}

The scatter in the plots could arise due to metallicity dependence on the parameters (Dotter et al. \citet{Dotter2008}; Feiden et al. \citet{Feiden2011}) or due to the contribution from unresolved binaries (Newton et al. \citet{Newton2015}).

\subsection{Stellar Metallicity Calibration}

\begin{figure*}
\centering
\begin{tabular}{c}
\hspace*{-0.45cm} \includegraphics[width= 7.25 in, height = 2.5 in, clip]{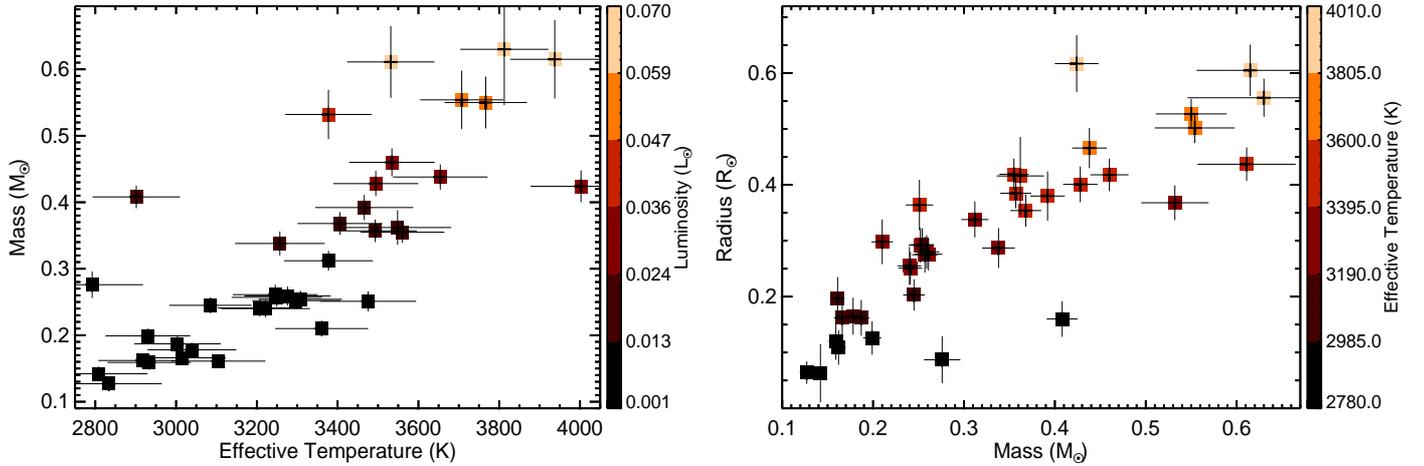}
\end{tabular}
\caption{Left: Stellar Mass vs. Stellar $T_{eff}$ where the points are shaded with colors according to the luminosity. Right: stellar radius vs. stellar mass  where the points are shaded with colors according to the stellar $T_{eff}$.}
\label{fig:fig11}
\end{figure*}

\begin{table}
 \centering
 \caption{M Dwarf Metallicity calibration samples}
 \label{tab:table7_metallicity_cal}
 \begin{tabular}{lccc}
  \hline \hline
Star & Sp. Type & \underline{SPOCS I} & \underline{Predicted}  \\
   &  & [Fe/H] \hspace*{0.5cm} [M/H] & [Fe/H] \hspace*{0.5cm} [M/H]  \\
 \hline
HD 79210  & M0 & -0.11 \hspace*{0.5cm}   -0.11 & -0.03 \hspace*{0.5cm}   0.10  \\
HD 28343   & M0 & 0.20 \hspace*{0.5cm}    0.01 &  0.22 \hspace*{0.5cm}  -0.06   \\
HD 233153    & M1.4 & 0.19 \hspace*{0.5cm}  0.16 &  0.21 \hspace*{0.5cm}   0.20   \\
HD 50281B  & M1.6 & 0.14 \hspace*{0.5cm}   -0.01 &  0.13 \hspace*{0.5cm}  -0.08  \\
Gl 797B   & M1.8 & -0.09 \hspace*{0.5cm}   -0.09 &  0.05 \hspace*{0.5cm}   0.20   \\
HD 119850   & M1.7 & -0.63 \hspace*{0.5cm} -0.68 & -0.48 \hspace*{0.5cm}  -0.37   \\
GJ 251   & M1.7 & -0.30 \hspace*{0.5cm} -0.39 & -0.22 \hspace*{0.5cm}  -0.14   \\
Gl 166C     & M4.8 & -0.28 \hspace*{0.5cm} -0.08 & -0.19 \hspace*{0.5cm}   0.16   \\
Gl 324B  & M2.6 & 0.31 \hspace*{0.6cm}      0.25 & -0.03 \hspace*{0.5cm}  -0.03   \\
Gl 406   & M4.8 & 0.25 \hspace*{0.8cm}       -   &  0.41 \hspace*{0.8cm}    -     \\
GJ 402     & M2.8 & 0.16 \hspace*{0.8cm}   -   & -0.02 \hspace*{0.8cm}    -     \\
Gl 905     & M4.8 & 0.23 \hspace*{0.8cm}     -   & -0.03 \hspace*{0.8cm}    -     \\
  
  \hline
\end{tabular}
 \vspace{1ex} 
\end{table}

We estimated the metallicity values [Fe/H] and [M/H] using the linear relations defined by Rojas-Ayala et al. (\citet{RojasAyala2012}). Rojas-Ayala et al. presented an updated K-band metallicity calibration relations using the Na I (2.206 $\mu$m) and Ca I (2.261 $\mu$m) features where the strength of these features are weighted by the $H_{2}O$-K2 index to remove the $T_{eff}$ dependence of the features. We choose 12 dwarfs as [Fe/H] calibrators and 9 for [M/H] calibration from the SPOCS I catalogue and established the best-fit relationship by performing multivariate linear regression. The calibration sample is given in Table \ref{tab:table7_metallicity_cal} and the relation for [Fe/H] is:
\begin{equation}
\label{eqn56}
[Fe/H] = -0.571 + 0.094\times {\frac{NaI_{EW}}{H_{2}O-K2}} + 0.007\times {\frac{CaI_{EW}}{H_{2}O-K2}} 
\end{equation}
\hspace*{0.8cm} RMSE ([Fe/H]) = 0.257 \\
\hspace*{0.92cm} MAD ([Fe/H]) = 0.13 \\

\hspace*{-0.7 cm} And the calibration relation for total metallicity [M/H] is:
\begin{equation}
\label{eqn58}
[M/H] = -1.143 + 0.096\times {\frac{NaI_{EW}}{H_{2}O-K2}} + 0.139\times {\frac{CaI_{EW}}{H_{2}O-K2}}
\end{equation}
\hspace*{0.8cm} RMSE ([M/H]) = 0.217 \\
\hspace*{0.92cm} MAD ([M/H]) = 0.15 \\

\hspace*{-0.7 cm} We have calculated the RMSE and the MAD for both of our metallicity relations. The RMSE indicates the accuracy of the response of the regression models where lower the value means better the fits. The RMSE values for both of our fits are RMSE(Fe/H) = 0.257 and RMSE (M/H) = 0.217. Now to estimate the uncertainties in metallicities using the calibration relations, we used a MCMC simulation code implemented with\texttt{ emcee}\footnote{\url{https://github.com/dfm/emcee/tree/master/emcee}} (Foreman-Mackey et al. \citet{Foreman2013}). First, we assumed the true value is normally distributed around the solution obtained from the least square fits with standard deviation equal to the intrinsic scatter of the fits. Uncertainties in the coefficients are calculated using maximum likelihood where we applied 500 walkers and 100 steps with random seed parameters. Finally, we measured the errors in metallicities using \href{https://github.com/mcfit/idl_emcee}{\texttt{idl\textunderscore emcee}}\footnote{\url{https://github.com/mcfit/idl_emcee}} by combining the errors of the coefficients with the errors from the features and indices with a $3\sigma$ confidence interval. \\

\begin{figure}
\centering
\begin{tabular}{c}
\hspace*{-0.55cm} \includegraphics[width= 3.45 in, height = 2.5 in, clip]{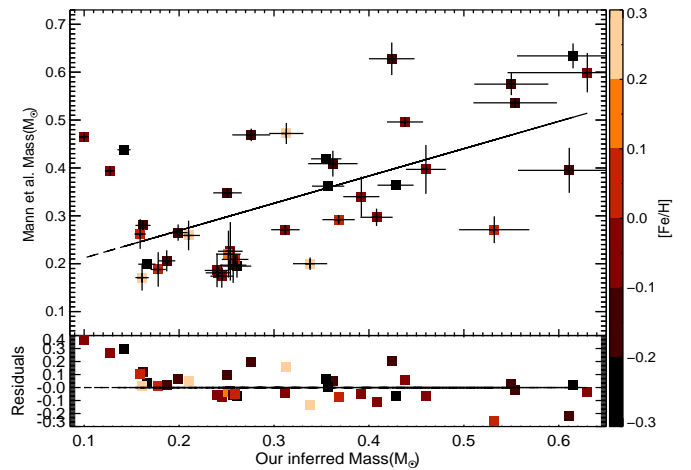}
\end{tabular}
\caption{Comparison of our inferred Mass with Mass estimated using the relation given by Mann et al. (\citet{Mann2015}). The points are shaded with colors according to the metallicity.}
\label{fig:fig12}
\end{figure}

$~~~~~~~~~$ EW of the Ca I vs EW of the Na I, both weighted by the $H_{2}O$ index for our observed M dwarfs are plotted in Figure \ref{fig:fig10}, where the isometallicity contours for both [Fe/H] and [M/H] calculated from the equations (\ref{eqn56}) and (\ref{eqn58}) are also shown. Stars with the maximum [Fe/H] and [M/H] values are Gl 324B (+0.33 dex) and V*CUCnc (+0.40 dex) and the  respective minimum values are GJ 661 (- 0.27 dex) and Gl 406 (- 0.59 dex). We have calculated the metallicity of 18 M dwarfs which were not present in Rojas-Ayala et al. (\citet{RojasAyala2012}) and 13 new objects that were not present in Mann et al.'s (\citet{Mann2015}) work (results are shown in Table \ref{tab:table9_finalresult}).

\subsection{Luminosity-Mass relation} 
Mann et al.\citet{Mann2013a} established a ($T_{eff}-M_{\star}$) relation by expressing the Mass of late-type K and M dwarf stars as a function of $T_{eff}$. Later  Mann et al. (\citet{Mann2015}) gave a semi-empirical $M_{K_{s}}$-Mass relation using masses derived from stellar models and observed absolute $K_{s}$ band luminosities. Motivated by this work we also estimated the Mass for our dwarf stars using a fourth order polynomial as a function of empirically estimated luminosity $(L/L{\sun})$ :
\begin{equation} 
\begin{split}
\label{eqn332}
M/M_{\odot} &=\ 0.0899 + 24.704\times (L/L_{\odot}) - 694.63\times (L/L_{\odot})^{2} \\
            &\ \ \ \ + 11202.4\times (L/L_{\odot})^{3} - 67773.4\times (L/L_{\odot})^{4} \\
\end{split}
\end{equation}
The best fit relation is obtained by using photometrically determined parameters of 8 calibration samples from Mann et al.\citet{Mann2013a} and parameters of 4 calibrators estimated using semi-empirical relation taken from Mann et al. (\citet{Mann2015}) (see the calibration sample in Table \ref{tab:table8_mass_cal}). The uncertainty in mass is calculated from the error in luminosity using \href{https://github.com/mcfit/idl_emcee}{\texttt{idl\textunderscore emcee}}. In the left side of Figure \ref{fig:fig11} we show the plot of Stellar Mass vs. Stellar $T_{eff}$ where the points are color-coded by the luminosity and in the right side we show stellar radius vs. stellar mass where the points are shaded with colors according to the $T_{eff}$ value. In Figure \ref{fig:fig12}, we show the comparison of our inferred Mass with Mass estimated using
relation given by Mann et al. (\citet{Mann2015}). The median and standard deviation of the differences in our calculated mass values with mass given by Mann et al. (\citet{Mann2015})  are respectively $0.06M_{\odot}$ and $0.09_{\odot}$.

\begin{table}
 \centering
 \caption{Mass calibration samples}
 \label{tab:table8_mass_cal}
 \begin{tabular}{lccccc}
  \hline
Star & $M_{\star}$ & $\sigma{M_{\star}}$ & $L_{\star}$ & $\sigma{L_{\star}}$ & Ref. \\
   & ($M_{\sun}$) &  & ($L_{\sun}$) &  & \\
 \hline
V*GXAnd  & 0.405 & 0.041 & 0.02256 & 0.00027 & a \\
HD 36395   & 0.637 & 0.064 & 0.06449 & 0.00139 & a \\
HD 119850    & 0.490 & 0.049 & 0.03694 & 0.00051 & a \\
HD 95735  & 0.392 & 0.039 & 0.02134 & 0.00030 & a \\
HD 79210   & 0.630 & 0.063 & 0.07316 & 0.00276 & a \\
GJ 338B   & 0.617 & 0.062 & 0.06875  & 0.00229 & a \\
GJ 436    & 0.447	 & 0.045 & 0.02834 & 0.00072 & a \\
GJ 581  & 0.308	 & 0.031 & 0.01181 & 0.00021 & a \\
GL 905   & 0.145	 & 0.015 & 0.00237 & 0.00010 & b \\
GL 447    & 0.168 & 0.017 & 0.00337 & 0.00020 & b \\
GL 166C     & 0.222	 & 0.022 & 0.00678 & 0.00010 & b \\
GJ 3379   & 0.226		& 0.023 & 0.00662 & 0.01200 & b \\
  
  \hline
\end{tabular}
 \vspace{1ex} 

     \raggedright \small $^{a}$ Mann et al.\citet{Mann2013a}.\\
     \raggedright $^{b}$ Mann et al. (\citet{Mann2015}).
\end{table} 

\section{Summary and Conclusion}
$~~~~$ In this paper, we present H (1.50-1.84 $\mu$m) and K (1.95-2.45 $\mu$m) band spectra of 53 M dwarfs including  seven stars with interferometric radius, nine objects with reliable metallicity estimates (SPOCS I catalogue), ten possible planet candidates and almost half of our sample is in the northern 8 pc region. We measured the EWs and $H_{2}O$ index in both H and K-band and determined spectroscopic $T_{eff}$, radius, luminosity, spectral type, absolute $K_{s}$ magnitude, metallicity, and mass, by deriving empirical calibration relations for these parameters. The estimation of our spectral properties is mostly model-independent.
 
\begin{enumerate}
\item[1.] Using H-band EWs of Mg(1.57 $\mu$m), Al (1.67 $\mu$m), Mg (1.71 $\mu$m) features and the $H_{2}O$-H index, we performed linear regression on 12 M dwarf calibrator stars with interferometrically and semi-empirically measured $T_{eff}$, radius and luminosity to calibrate our new relationships. Our calibrators lie within the parameter space of 2930 < $T_{eff}$(K) < 3930, 0.189 < $R/R_{\sun}$ < 0.608 and \textendash 2.44 < log $L/L_{\sun}$ < \textendash 1.10. Statistical tests show the standard deviations in the residuals of our best fits are 102K, 0.027$R_{\sun}$, and 0.12dex, respectively. These model-independent determinations can be applied to stars without parallax distances.
\item[2.] We took both H and K-band $H_{2}O$ indices and derive calibration relations for NIR spectral type and absolute magnitude as a linear function of the indices. Previously Rojas-Ayala et al. (\citet{RojasAyala2012}) used only the K-band water index and Newton at al. (\citet{Newton2014}) used only the H-band water index to calibrate sp.type and absolute $K_{s}$ magnitude respectively. For completeness we used both the indices and predicted the parameters with an accuracy of 0.7 subtypes and 0.6 ($M_{k}$) error within $3\sigma$ confidence interval and the higher adjusted square of the multiple correlation coefficient values ($R^{2}_{ap}$(sp.type) = 0.90 and $R^{2}_{ap}$($M_{k}$) = 0.83) also supports that the quality of these fits are good statistically. We show the existence of relationships between our inferred $T_{eff}$, radii, and luminosity with star's absolute $K_{s}$ magnitude, which in turn independently confirms the validation of our calibrations. We also estimated the spectroscopic distances of our stars using $M_{K_{s}}$ and photometrically determined apparent magnitudes.
\item[3.] We compared our results to the literature, to identify the most suitable calibration relation for overall metallicity [M/H] and iron abundance [Fe/H], and we found the best results using the relations given by Rojas-Ayala et al. (\citet{RojasAyala2012}). The metallicity dependence of K-band Na I and Ca I features for our observed M dwarf spectra motivated us to use them for calibration. We estimated the metallicity of 18 new M dwarfs that were not analyzed by Rojas-Ayala et al. (\citet{RojasAyala2012}) in their work. We found that the metallicity values of our objects lie in both sub-solar and super-solar regions, confirming young as well as old dwarf stars in our observed sample, with the probably planet hosts lying in the solar metallicity region.
\item[4.] We used stars with photometrically estimated mass and luminosity (L/$L_{\sun}$) as calibrators and determined the mass of the observed sample empirically as a fourth-order polynomial of luminosity. 

\end{enumerate}

\section{Acknowledgements}
The authors are thankful to Prof. David James Pinfield for all the valuable comments and suggestions, which helped to improve the overall quality of this paper. This research work is supported by S. N. Bose National Centre for Basic Sciences under the Department of Science and Technology (DST), Govt. of India. DK gratefully acknowledges the DST, Govt. of India for the INSPIRE Fellowship scheme. The authors are thankful to IIA, Bangalore for providing observing time in 2-m HCT (IIA) telescope for the present work. The authors also acknowledge the facilities at IAO, CREST and the staff of IAO, Hanle who made these observations possible and the usage of the TIRSPEC instrument. \\

\begin{table*} 
 \centering 
 \caption{Estimated Stellar parameters for our observed (2-m HCT) M-Dwarfs:}
 \label{tab:table9_finalresult}
 \begin{tabular}{lccccccccc}
  \hline \hline
Name & Sp.T. $^{a}$& $T_{eff}$ $^{b}$& Radius $^{b}$& Luminosity $^{b}$& Mass $^{c}$& $M_{k_{s}}$ $^{d}$& Dist. $^{e}$& [M/H] $^{f}$& [Fe/H] $^{f}$\\
 & & (K) & $R_{\sun}$ & log $L/L_{\sun}$ & $M_{\sun}$ & (mag) & (pc) & &\\
 \hline
HD232979 	&	K7.4 	&3937 $\pm$	110 	&0.605 $\pm$	0.046 	&\textendash 1.217 $\pm$	0.133 	&0.615 $\pm$	0.059 	&3.837	$\pm$ 0.678  &17.458	$\pm$ 2.08 &\textendash 0.35 $\pm$	0.15 &	\textendash 0.21 	$\pm$ 0.07 \\	 
HD79210  	&	K7.8 	&4002  $\pm$	125 	&0.617 $\pm$	0.051 	&\textendash 1.601 $\pm$	0.160 	&0.424 $\pm$	0.024 	&4.549	$\pm$ 0.641  &7.881	$\pm$ 0.91 &\textendash 0.20 $\pm$	0.17 &	\textendash 0.08 	$\pm$ 0.09 \\
Gl338B 	 	&	M0.9 	&3844 $\pm$	127 	&0.582 $\pm$	0.047 	&\textendash 1.045 $\pm$	0.126 	&\textemdash       &5.133	$\pm$ 0.642  &6.330	$\pm$ 0.68 &\textendash 0.13 $\pm$ 0.16 &	\textendash 0.07 	$\pm$ 0.09 \\
HD28343  	&	K7.6 	&4073 $\pm$	112 	&0.706 $\pm$	0.042 	&\textendash 0.655 $\pm$	0.136 	&\textemdash       &4.065	$\pm$ 0.681  &14.521	$\pm$ 1.75 &0.08  $\pm$ 0.20 &	\textendash 0.02 	$\pm$ 0.10 \\
GJ3395 		&	M1.5 	&3867 $\pm$	113 	&0.584 $\pm$	0.040 	&\textendash 1.023 $\pm$	0.127 	&\textemdash       &5.639	$\pm$ 0.632  &19.266	$\pm$ 2.16 &\textendash 0.06 $\pm$ 0.16 &	\textendash 0.03 	$\pm$ 0.08 \\
HD233153 	&	M1.4 	&3812 $\pm$	109 	&0.556 $\pm$	0.034 	&\textendash 1.172 $\pm$	0.137 	&0.630 $\pm$	0.084 	&5.957	$\pm$ 0.802  &9.129	$\pm$ 1.30 &\textendash 0.03 $\pm$	0.25 &	\textendash 0.02 	$\pm$ 0.13 \\
GJ 514 	&	M0.7 	&2901 $\pm$	108 	&0.160 $\pm$	0.032 	&\textendash 1.643 $\pm$	0.124 	&0.408 $\pm$	0.017 	&5.098	$\pm$ 0.679  &10.975	$\pm$ 1.32 &\textendash 0.13 $\pm$	0.17 &	\textendash 0.06 	$\pm$ 0.10 \\
GJ 525 	&	M1.7 	&3548 $\pm$	133 	&0.416 $\pm$	0.070 	&\textendash 1.766 $\pm$	0.201 	&0.362 $\pm$	0.026 	&6.124	$\pm$ 1.148  &10.452	$\pm$ 1.96 &\textendash 0.52 $\pm$	0.20 &	\textendash 0.16 	$\pm$ 0.10 \\
HD36395 	&	M2.6 	&3849 $\pm$	275 	&0.553 $\pm$	0.149 	&\textendash 1.320 $\pm$	0.390 	&0.554 $\pm$	0.193 	&6.824	$\pm$ 1.343  &2.601	$\pm$ 0.62 &  \textemdash    &   \textemdash    \\
GJ 625 		&	M3.1 	&3221 $\pm$	110 	&0.255 $\pm$	0.034 	&\textendash 2.131 $\pm$	0.131 	&0.240 $\pm$	0.013 	&7.210	$\pm$ 0.649  &5.304	$\pm$ 0.61 &\textendash 0.08 $\pm$	0.18 &	0.02 	$\pm$ 0.09 \\
HD95735 	&	M2.1 	&3560 $\pm$	104 	&0.419 $\pm$	0.028 	&\textendash 1.784 $\pm$	0.126 	&0.355 $\pm$	0.016 	&6.179	$\pm$ 0.646  &2.705     $\pm$ 0.31 &\textendash 0.45 $\pm$	0.19 &	\textendash 0.18 	$\pm$ 0.13 \\
HD115953 	&	M1.2 	&3766 $\pm$	102 	&0.527 $\pm$	0.027 	&\textendash 1.327 $\pm$	0.127 	&0.550 $\pm$	0.039 	&5.484	$\pm$ 0.624  &6.339	$\pm$ 0.72 &\textendash 0.16 $\pm$	0.16 &	\textendash 0.09 	$\pm$ 0.08 \\
V*GXAnd 	&	M1.7 	&3493 $\pm$	103 	&0.385 $\pm$	0.027 	&\textendash 1.778 $\pm$	0.122 	&0.357 $\pm$	0.017 	&5.959	$\pm$ 0.631  &4.094	$\pm$ 0.44 &\textendash 0.40 $\pm$	0.14 &	\textendash 0.19 	$\pm$ 0.08 \\
V*BRPsc 	&	M1.9 	&3494 $\pm$	105 	&0.401 $\pm$	0.032 	&\textendash 1.591 $\pm$	0.125 	&0.428 $\pm$	0.019 	&5.773	$\pm$ 0.650  &7.145	$\pm$ 0.81 &\textendash 0.47 $\pm$	0.18 &	\textendash 0.27 	$\pm$ 0.09 \\
HD50281B 	&	M1.6 	&3465 $\pm$	121 	&0.380 $\pm$	0.044 	&\textendash 1.684 $\pm$	0.140 	&0.392 $\pm$	0.019 	&5.776	$\pm$ 0.842  &9.759	$\pm$ 1.40 &0.09  $\pm$ 0.20 &	0.01 	$\pm$ 0.10 \\
GJ 494 	&	M2.4 	&3377 $\pm$	107 	&0.368 $\pm$	0.031 	&\textendash 1.360 $\pm$	0.135 	&0.532 $\pm$	0.037 	&6.409	$\pm$ 0.683  &6.820	$\pm$ 0.83 &0.31  $\pm$ 0.18 &	0.09 	$\pm$ 0.10 \\
GJ 649 	&	M1.2 	&3654 $\pm$	117 	&0.466 $\pm$	0.036 	&\textendash 1.566 $\pm$	0.138 	&0.438 $\pm$	0.019 	&5.847	$\pm$ 0.657  &9.024	$\pm$ 1.06 &\textendash 0.06 $\pm$	0.21 &	\textendash 0.03 	$\pm$ 0.12 \\
GL436 		&	M1.9 	&3534 $\pm$	106 	&0.418 $\pm$	0.029 	&\textendash 1.511 $\pm$	0.126 	&0.460 $\pm$	0.021 	&5.848	$\pm$ 0.675  &11.092	$\pm$ 1.31 &\textendash 0.12 $\pm$	0.19 &	\textendash 0.06 	$\pm$ 0.08 \\
GL176 		&	M1.9 	&3377 $\pm$	110 	&0.338 $\pm$	0.032 	&\textendash 1.901 $\pm$	0.129 	&0.312 $\pm$	0.015 	&6.403	$\pm$ 0.865  &6.931	$\pm$ 1.05 &\textendash 0.13 $\pm$	0.21 &	\textendash 0.01 	$\pm$ 0.09 \\
GL797B 	&	M1.8 	&2792 $\pm$	126 	&0.087 $\pm$	0.042 	&\textendash 2.009 $\pm$	0.189 	&0.276 $\pm$	0.020 	&6.038	$\pm$ 0.641  &18.863	$\pm$ 2.20 &\textendash 0.29 $\pm$	0.20 &	\textendash 0.14 	$\pm$ 0.15 \\
HD119850 	&	M1.7 	&3707 $\pm$	103 	&0.502 $\pm$	0.027 	&\textendash 1.321 $\pm$	0.123 	&0.554 $\pm$	0.044 	&5.900	$\pm$ 0.637  &5.047	$\pm$ 0.57 &\textendash 0.31 $\pm$	0.16 &	\textendash 0.15 	$\pm$ 0.10 \\
GJ 628 	&	M4.0 	&3085 $\pm$	102 	&0.203 $\pm$	0.028 	&\textendash 2.112 $\pm$	0.123 	&0.245 $\pm$	0.012 	&7.685	$\pm$ 0.657  &3.006	$\pm$ 0.35 &\textendash 0.05 $\pm$	0.16 &	0.02 	$\pm$ 0.08 \\
GJ 251 	&	M1.7 	&2930 $\pm$	105 	&0.126 $\pm$	0.030 	&\textendash 2.295 $\pm$	0.131 	&0.199 $\pm$	0.010 	&6.024	$\pm$ 0.650  &7.099	$\pm$ 0.79 &\textendash 0.25 $\pm$	0.15 &	\textendash 0.08 	$\pm$ 0.07 \\
GJ 109 		&	M3.2 	&2916 $\pm$	109 	&0.109 $\pm$	0.031 	&\textendash 2.493 $\pm$	0.147 	&0.162 $\pm$	0.008 	&6.840	$\pm$ 0.632  &6.668	$\pm$ 0.77 &\textendash 0.22 $\pm$	0.18 &	\textendash 0.07 	$\pm$ 0.10 \\
GJ48 		&	M2.7 	&3531 $\pm$	108 	&0.437 $\pm$	0.030 	&\textendash 1.226 $\pm$	0.133 	&0.611 $\pm$	0.054 	&6.458	$\pm$ 0.668  &6.283	$\pm$ 0.72 &\textendash 0.08 $\pm$	0.18 &	\textendash 0.11 	$\pm$ 0.09 \\
Gl581 		&	M3.6 	&3475 $\pm$	119 	&0.364 $\pm$	0.045 	&\textendash 2.091 $\pm$	0.150 	&0.251 $\pm$	0.015 	&7.598	$\pm$ 0.681  &4.450	$\pm$ 0.54 &\textendash 0.11 $\pm$	0.20 &	\textendash 0.11 	$\pm$ 0.12 \\
GJ 661 	&	M3.2 	&3245 $\pm$	105 	&0.276 $\pm$	0.029 	&\textendash 2.058 $\pm$	0.136 	&0.261 $\pm$	0.015 	&7.139	$\pm$ 0.639  &3.453	$\pm$ 0.37 &\textendash 0.38 $\pm$	0.16 &	\textendash 0.27 	$\pm$ 0.06 \\
GJ 3378 	&	M3.1 	&3307 $\pm$	103 	&0.293 $\pm$	0.029 	&\textendash 2.082 $\pm$	0.124 	&0.254 $\pm$	0.013 	&6.789	$\pm$ 0.648  &9.336	$\pm$ 1.06 &\textendash 0.12 $\pm$	0.16 &	\textendash 0.01 	$\pm$ 0.08 \\
GJ 169.1 A 	&	M2.9 	&3297 $\pm$	109 	&0.292 $\pm$	0.031 	&\textendash 2.089 $\pm$	0.125 	&0.252 $\pm$	0.013 	&6.797	$\pm$ 0.649  &6.090	$\pm$ 0.71 &0.14  $\pm$ 0.16 &	0.21 	$\pm$ 0.09 \\
GJ 273 	&	M3.6 	&3249 $\pm$	112 	&0.275 $\pm$	0.032 	&\textendash 2.072 $\pm$	0.127 	&0.257 $\pm$	0.014 	&7.411	$\pm$ 0.668  &3.089	$\pm$ 0.36 &\textendash 0.24 $\pm$	0.29 &	\textendash 0.09 	$\pm$ 0.12 \\
GJ 3522 		&	M3.5 	&3207 $\pm$	105 	&0.251 $\pm$	0.029 	&\textendash 2.127 $\pm$	0.127 	&0.241 $\pm$	0.013 	&6.872	$\pm$ 0.632  &5.797	$\pm$ 0.64 &\textendash 0.16 $\pm$	0.15 &	\textendash 0.03 	$\pm$ 0.08 \\
GJ 3379 		&	M3.6 	&3276 $\pm$	107 	&0.280 $\pm$	0.030 	&\textendash 2.064 $\pm$	0.122 	&0.259 $\pm$	0.014 	&7.108	$\pm$ 0.649  &6.115	$\pm$ 0.68 &\textendash 0.06 $\pm$	0.17 &	0.09 	$\pm$ 0.08 \\
Gl447 		&	M4.1 	&3039 $\pm$	110 	&0.165 $\pm$	0.033 	&\textendash 2.399 $\pm$	0.134 	&0.178 $\pm$	0.008 	&7.570	$\pm$ 0.628  &4.130	$\pm$ 0.44 &\textendash 0.08 $\pm$	0.16 &	0.07 	$\pm$ 0.08 \\
G246-33 	&	M5.9 	&\textemdash 	&\textemdash 	&\textemdash  	 	&\textemdash	&10.11	$\pm$ 1.151  &5.129	$\pm$ 1.01 &\textemdash    &   \textemdash  \\ 
GJ213 		&	M2.6 	&2808 $\pm$	121 	&0.063 $\pm$	0.052 	&\textendash 2.644 $\pm$	0.179 	&0.142 $\pm$	0.007 	&6.094	$\pm$ 0.886  &11.46	$\pm$ 1.76 &\textendash 0.42 $\pm$	0.23 &	\textendash 0.24 	$\pm$ 0.16 \\
GJ445 		&	M4.1 	&3014 $\pm$	108 	&0.162 $\pm$	0.033 	&\textendash 2.466 $\pm$	0.159 	&0.166 $\pm$	0.009 	&7.622	$\pm$ 0.628  &4.639	$\pm$ 0.51 &\textendash 0.46 $\pm$	0.16 &	\textendash 0.18 	$\pm$ 0.06 \\
GJ 402 	&	M2.8 	&2388 $\pm$	113 	&\textemdash 	&\textendash 2.542 $\pm$	0.130 	&0.155 $\pm$	0.007 	&6.695	$\pm$ 0.640  &8.610	$\pm$ 0.97 &\textendash 0.01 $\pm$	0.17 &	0.18 	$\pm$ 0.10 \\
GJ 408 	&	M2.5 	&3360 $\pm$	115 	&0.298 $\pm$	0.040 	&\textendash 2.246 $\pm$	0.141 	&0.210 $\pm$	0.012 	&6.688	$\pm$ 0.654  &5.786	$\pm$ 0.64 &0.28  $\pm$ 0.28 &	0.25 	$\pm$ 0.16 \\
GJ 277 B 	&	M3.3 	&3405 $\pm$	105 	&0.354 $\pm$	0.029 	&\textendash 1.748 $\pm$	0.123 	&0.368 $\pm$	0.017 	&6.952	$\pm$ 0.631  &9.154	$\pm$ 1.06 &0.12  $\pm$ 0.18 &	0.07 	$\pm$ 0.10 \\
GJ 873 	&	M2.6 	&2518 $\pm$	106 	&\textemdash 	&\textendash 2.457 $\pm$	0.124 	&0.168 $\pm$	0.007 	&6.525	$\pm$ 0.670  &5.662	$\pm$ 0.67 &\textendash 0.06 $\pm$	0.15 &	0.02 	$\pm$ 0.07 \\
GJ 1156 		&	M2.9 	&\textemdash 	&\textemdash 	&\textendash 2.851 $\pm$	0.370 	&0.123 $\pm$	0.008 	&7.518	$\pm$ 1.715  &10.24	$\pm$ 2.90 &\textemdash    &   \textemdash    \\
GJ 3454 		&	M4.5 	&2680 $\pm$	140 	&\textemdash 	&\textendash 2.638 $\pm$	0.133 	&0.143 $\pm$	0.006 	&7.815	$\pm$ 0.623  &7.816	$\pm$ 0.87 &\textendash 0.03 $\pm$	0.22 &	0.11 	$\pm$ 0.13 \\
GJ 166C 		&	M4.8 	&3003 $\pm$	107 	&0.162 $\pm$	0.032 	&\textendash 2.353 $\pm$	0.129 	&0.187 $\pm$	0.009 	&7.984	$\pm$ 0.656  &3.941	$\pm$ 0.45 &\textendash 0.24 $\pm$	0.16 &	\textendash 0.09 	$\pm$ 0.07 \\
GJ 3304 		&	M3.9 	&2794 $\pm$	146 	&\textemdash 	&\textendash 3.351 $\pm$	0.244 	&0.100 $\pm$	0.002 	&7.237	$\pm$ 0.751  &10.438	$\pm$ 1.30 &\textendash 0.34 $\pm$	0.28 &	\textendash 0.05 	$\pm$ 0.13 \\
GJ 324B 		&	M2.6 	&3623 $\pm$	124 	&\textemdash 	&\textendash 1.899 $\pm$	0.145 	&0.313 $\pm$	0.018 	&6.316	$\pm$ 0.635  &18.655	$\pm$ 2.11 &0.29  $\pm$ 0.21 &	0.33 	$\pm$ 0.14 \\
GJ 3421 		&	M4.9 	&2648 $\pm$	109 	&\textemdash 	&\textendash 2.778 $\pm$	0.151 	&0.129 $\pm$	0.005 	&7.613	$\pm$ 0.637  &10.75	$\pm$ 1.19 &\textendash 0.57 $\pm$	0.15 &	\textendash 0.19 	$\pm$ 0.09 \\
GJ 905 		&	M4.8 	&3104 $\pm$	117 	&0.197 $\pm$	0.038 	&\textendash 2.506 $\pm$	0.130 	&0.161 $\pm$	0.007 	&8.482	$\pm$ 0.633  &3.086	$\pm$ 0.33 &0.13  $\pm$ 0.18 &	0.26 	$\pm$ 0.11 \\
V*CUCnc 	&	M3.3 	&3256 $\pm$	111 	&0.287 $\pm$	0.036 	&\textendash 1.830 $\pm$	0.133 	&0.338 $\pm$	0.018 	&6.766	$\pm$ 0.648  &9.264	$\pm$ 1.03 &0.40  $\pm$ 0.21 &	0.25 	$\pm$ 0.10 \\
GJ 1286 		&	M4.2 	&\textemdash 	&\textemdash 	&\textendash 1.335 $\pm$	0.400 	&0.546 $\pm$	0.180 	&7.286	$\pm$ 0.960  &15.09	$\pm$ 2.42 &\textendash 0.18 $\pm$	0.16 &	0.07 	$\pm$ 0.08 \\
GJ 473 	&	M4.9 	&\textemdash 	&\textemdash 	&\textendash 2.997 $\pm$	0.127 	&0.114 $\pm$	0.003 	&8.191	$\pm$ 0.648  &3.714	$\pm$ 0.42 &\textendash 0.04 $\pm$	0.18 &	0.10 	$\pm$ 0.07 \\
GJ 406 		&	M4.8 	&\textemdash 	&\textemdash 	&\textendash 2.907 $\pm$	0.152 	&0.119 $\pm$	0.003 	&9.161	$\pm$ 0.673  &2.424	$\pm$ 0.29 &\textendash 0.59 $\pm$	0.26 &	\textendash 0.17 	$\pm$ 0.21 \\
GJ 1083 	&	M5.4 	&2833 $\pm$	131 	&\textemdash 	&\textendash 2.789 $\pm$	0.171 	&0.127 $\pm$	0.005 	&8.499	$\pm$ 0.662  &8.091	$\pm$ 0.92 &\textendash 0.02 $\pm$	0.25 &	0.02 	$\pm$ 0.08 \\
GAT1370 	&	M7.6 	&\textemdash 	&\textemdash 	&\textendash 0.887 $\pm$	0.770 	&\textemdash 	&9.038	$\pm$ 1.029  &5.122	$\pm$ 0.89 &\textemdash    &   \textemdash    \\
  
  \hline
\end{tabular}
 \vspace{1ex} 
 
\hspace*{-17.cm} \textbf{Notes.}

          \raggedright \small $^{a}$ Spectral types are calculated using equation (\ref{eqn34}) having errors of the order of roughly $\pm$0.7 subtypes. \\
          \raggedright \small $^{b}$ $T_{eff}$, radius and luminosity are estimated using equations (\ref{eqn22}), (\ref{eqn23}) and (\ref{eqn24}) respectively. \\
          \raggedright \small $^{c}$ Stellar mass is measured using mass-luminosity equation (\ref{eqn332}). \\
          \raggedright \small $^{d}$ $M_{K_{s}}$ is estimated using equation (\ref{eqn111}). \\
          \raggedright \small $^{e}$ Spectroscopic distance is calculated using estimated absolute K magnitude and photometrically determined apparent magnitude while we ignored the extinction effects. \\
          \raggedright \small $^{f}$ [Fe/H] and [M/H] are calculated using equation (\ref{eqn56}) and (\ref{eqn58}) respectively.          
\end{table*}


\begin{table}
 \centering
 \caption{EW of Mg I (1.57 $\mu$m) for the BT-Settl model spectra (resolution downgraded to that of the TIRSPEC instrument)}.
 \label{tab:table10_mg157ew}
 \begin{tabular}{lccccc}
  \hline \hline
  \textbf{$T_{eff}$} & \textbf{log g = 4.5} & &  &\textbf{log g = 5.0} & \vspace{0.2cm}  \\

   & [Fe/H] = 0.0 & & [Fe/H] = - 0.5 & [Fe/H] = 0.0 & [Fe/H] = 0.5 \\
 \hline
  
2600& 0.425 &&  \textemdash   &  \textemdash   &  \textemdash \\
2700& 1.099 && \textemdash   &  \textemdash   &  \textemdash   \\
2800&   \textemdash   &&0.203 &0.693 &0.136 \\
2900&   \textemdash   &&0.911 &  \textemdash   &  \textemdash   \\
3000& 1.022 &&0.632 &  \textemdash   &1.188 \\
3100& 0.552 &&1.195 &0.624 &0.310 \\
3200& 0.928 &&0.166 &1.365 &1.315 \\
3300& 1.344 &&1.215 &1.450 &1.773 \\
3400& 1.948 &&1.334 &1.269 &2.880 \\
3500& 2.967 &&1.973 &1.885 &3.468 \\
3600& 3.479 &&1.836 &2.581 &3.893 \\
3700& 3.492 &&2.262 &2.806 &4.345 \\
3800& 4.209 &&3.292 &3.322 &5.469 \\
3900& 5.102 &&3.372 &4.036 &5.667 \\
4000& 5.857 &&3.856 &4.547 &6.340 \\
4100& 6.174 &&4.583 &5.787 &7.202 \\
  
  \hline
\end{tabular}
 \vspace{1ex} 
\end{table}

\begin{table}
 \centering
 \caption{EW of Al-a (1.67 $\mu$m) for the BT-Settl model spectra (resolution downgraded to that of the TIRSPEC instrument)}.
 \label{tab:table11_ala167ew}
 \begin{tabular}{lccccc}
  \hline \hline
  \textbf{$T_{eff}$} & \textbf{log g = 4.5} & &  &\textbf{log g = 5.0} & \vspace{0.2cm}  \\

   & [Fe/H] = 0.0 & & [Fe/H] = - 0.5 & [Fe/H] = 0.0 & [Fe/H] = 0.5 \\
 \hline
  
2600& 0.360 & &0.133& 0.088& 0.363 \\
2700& 0.312 & &0.136& 0.349& 0.488 \\
2800& 0.158 & &0.254& 0.444& 0.654 \\
2900& 0.501 & &0.335& 0.489& 0.954 \\
3000& 0.875 & &0.314& 0.640& 0.949 \\
3100& 1.099 & &0.448& 0.808& 1.100 \\
3200& 1.131 & &0.733& 1.228& 1.356 \\
3300& 1.299 & &0.726& 1.480& 1.619 \\
3400& 1.607 & &0.916& 1.505& 1.867 \\
3500& 1.600 & &0.973& 2.060& 2.079 \\
3600& 1.922 & &1.165& 1.569& 2.282 \\
3700& 1.700 & &1.177& 1.834& 2.378 \\
3800& 1.984 & &1.177& 1.623& 2.332 \\
3900& 1.687 & &1.073& 1.666& 2.806 \\
4000& 1.916 & &1.421& 1.996& 2.682 \\
4100& 1.975 & &1.070& 1.798& 2.706 \\
  
  \hline
\end{tabular}
 \vspace{1ex} 
\end{table}

\begin{table}
 \centering
 \caption{EW of Mg I (1.71 $\mu$m) for the BT-Settl model spectra (resolution downgraded to that of the TIRSPEC instrument)}.
 \label{tab:table12_mg171ew}
 \begin{tabular}{lccccc}
  \hline \hline
  \textbf{$T_{eff}$} & \textbf{log g = 4.5} & &  &\textbf{log g = 5.0} & \vspace{0.2cm}  \\

   & [Fe/H] = 0.0 & & [Fe/H] = - 0.5 & [Fe/H] = 0.0 & [Fe/H] = 0.5 \\
 \hline
  
2600& 0.224 & &  \textemdash  & 0.206&   \textemdash \\
2700&   \textemdash   & &  \textemdash  &  \textemdash   &   \textemdash  \\
2800& 0.362 & &0.253&  \textemdash   & 0.208 \\
2900& 0.395 & &0.442&  \textemdash   & 0.748 \\
3000& 0.406 & &0.535& 0.905& 1.130 \\
3100& 1.126 & &0.276& 0.903& 0.710 \\
3200& 1.294 & &1.012& 1.806& 2.160 \\
3300& 1.565 & &1.286& 1.759& 2.019 \\
3400& 2.472 & &1.122& 2.640& 2.913 \\
3500& 2.959 & &1.853& 2.165& 3.253 \\
3600& 2.820 & &2.114& 2.561& 3.364 \\
3700& 3.697 & &2.386& 3.237& 3.590 \\
3800& 3.580 & &2.504& 3.234& 4.366 \\
3900& 4.362 & &2.964& 3.528& 5.038 \\
4000& 4.154 & &3.565& 3.675& 5.301 \\
4100& 4.553 & &3.521& 4.380& 5.584 \\
  
  \hline
\end{tabular}
 \vspace{1ex} 
\end{table}

\begin{table}
 \centering
 \caption{EW of Na I (2.21 $\mu$m) for the BT-Settl model spectra (resolution downgraded to that of the TIRSPEC instrument)}.
 \label{tab:table13_na221_ew}
 \begin{tabular}{lccccc}
  \hline \hline
  \textbf{$T_{eff}$} & \textbf{log g = 4.5} & &  &\textbf{log g = 5.0} & \vspace{0.2cm}  \\

   & [Fe/H] = 0.0 & & [Fe/H] = - 0.5 & [Fe/H] = 0.0 & [Fe/H] = 0.5 \\
 \hline
  
2600& 2.273 & &1.883&  \textemdash   &2.693 \\
2700& 4.441 & &2.419&2.613 &3.566 \\
2800& 3.804 & &2.746&  \textemdash   &3.593 \\
2900& 3.691 & &2.390&  \textemdash   &3.055 \\
3000& 3.136 & &3.189&4.949 &3.638 \\
3100& 2.806 & &2.889&5.300 &4.166 \\
3200& 3.507 & &2.446&2.893 &4.382 \\
3300& 3.477 & &2.721&4.459 &4.077 \\
3400& 3.589 & &2.684&3.955 &4.761 \\
3500& 3.053 & &2.554&4.146 &4.507 \\
3600& 3.498 & &2.392&3.802 &4.601 \\
3700& 3.190 & &1.904&3.455 &4.380 \\
3800& 2.708 & &1.899&3.154 &4.667 \\
3900& 3.181 & &1.795&2.838 &4.743 \\
4000& 2.882 & &1.829&3.053 &4.907 \\ 
4100& 3.303 & &1.800&3.056 &4.897 \\
  
  \hline
\end{tabular}
 \vspace{1ex} 
\end{table}

\begin{table}
 \centering
 \caption{EW of Ca I (2.26 $\mu$m) for the BT-Settl model spectra (resolution downgraded to that of the TIRSPEC instrument)}.
 \label{tab:table14_ca226_ew}
 \begin{tabular}{lccccc}
  \hline \hline
  \textbf{$T_{eff}$} & \textbf{log g = 4.5} & &  &\textbf{log g = 5.0} & \vspace{0.2cm}  \\

   & [Fe/H] = 0.0 & & [Fe/H] = - 0.5 & [Fe/H] = 0.0 & [Fe/H] = 0.5 \\
 \hline
  
2600& 0.801 & &0.659&  \textemdash   &0.929 \\
2700& 0.790 & &0.570&0.115 &0.597 \\
2800&   \textemdash   & &0.425&  \textemdash   &0.724 \\
2900& 0.014 & &0.457&  \textemdash   &0.816 \\
3000& 0.905 & &0.618&0.679 &1.295 \\
3100& 1.003 & &0.493&0.805 &1.499 \\
3200& 1.434 & &0.506&0.941 &2.178 \\
3300& 1.826 & &0.696&1.330 &2.330 \\
3400& 1.941 & &0.933&2.010 &3.060 \\
3500& 1.862 & &0.977&1.695 &3.221 \\
3600& 2.164 & &0.971&1.848 &3.513 \\
3700& 2.179 & &0.857&1.853 &3.715 \\
3800& 2.112 & &0.955&1.863 &3.958 \\
3900& 2.304 & &0.969&1.822 &4.085 \\
4000& 2.280 & &0.923&1.900 &4.225 \\
4100& 2.324 & &1.003&1.980 &4.177 \\
  
  \hline
\end{tabular}
 \vspace{1ex} 
\end{table}

\begin{table}
 \centering
 \caption{H-band $H_{2}O$ index for the BT-Settl model spectra (resolution downgraded to that of the TIRSPEC instrument)}.
 \label{tab:table15_hband_waterindex}
 \begin{tabular}{lccccc}
  \hline \hline
  \textbf{$T_{eff}$} & \textbf{log g = 4.5} & &  &\textbf{log g = 5.0} & \vspace{0.2cm}  \\

   & [Fe/H] = 0.0 & & [Fe/H] = - 0.5 & [Fe/H] = 0.0 & [Fe/H] = 0.5 \\
 \hline
  
2600&0.611  & &0.597&0.604 &0.605 \\
2700&0.683  & &0.645&0.688 &0.648 \\
2800&0.670  & &0.680&0.709 &0.686 \\
2900&0.709  & &0.715&0.735 &0.736 \\
3000&0.762  & &0.736&0.787 &0.748 \\
3100&0.774  & &0.778&0.811 &0.786 \\
3200&0.817  & &0.802&0.825 &0.819 \\
3300&0.844  & &0.828&0.847 &0.844 \\
3400&0.873  & &0.847&0.890 &0.869 \\
3500&0.895  & &0.869&0.908 &0.896 \\
3600&0.926  & &0.896&0.928 &0.925 \\
3700&0.926  & &0.933&0.944 &0.938 \\
3800&0.942  & &0.947&0.958 &0.946 \\
3900&0.944  & &0.959&0.960 &0.937 \\
4000&0.943  & &0.970&0.954 &0.942 \\
4100&0.940  & &0.969&0.955 &0.933 \\
  
  \hline
\end{tabular}
 \vspace{1ex} 
\end{table}

\begin{table}
 \centering
 \caption{K-band $H_{2}O$ index for the BT-Settl model spectra (resolution downgraded to that of the TIRSPEC instrument)}.
 \label{tab:table16_kband_waterindex}
 \begin{tabular}{lccccc}
  \hline \hline
  \textbf{$T_{eff}$} & \textbf{log g = 4.5} & &  &\textbf{log g = 5.0} & \vspace{0.2cm}  \\

   & [Fe/H] = 0.0 & & [Fe/H] = - 0.5 & [Fe/H] = 0.0 & [Fe/H] = 0.5 \\
 \hline
  
2600&0.691  & &0.728&  \textemdash   &0.663 \\
2700&0.745  & &0.755&0.699 &0.705 \\
2800&0.756  & &0.773&  \textemdash   &0.739 \\
2900&0.789  & &0.801&  \textemdash   &0.773 \\
3000&0.814  & &0.821&0.832 &0.799 \\
3100&0.834  & &0.844&0.858 &0.828 \\
3200&0.861  & &0.855&0.844 &0.849 \\
3300&0.882  & &0.888&0.903 &0.869 \\
3400&0.907  & &0.908&0.927 &0.903 \\
3500&0.927  & &0.928&0.956 &0.921 \\
3600&0.957  & &0.952&0.960 &0.950 \\
3700&0.974  & &0.980&0.980 &0.984 \\
3800&1.006  & &0.998&1.006 &0.993 \\
3900&1.005  & &1.019&1.025 &1.001 \\
4000&1.026  & &1.030&1.028 &1.002 \\
4100&1.011  & &1.043&1.033 &1.006 \\
  
  \hline
\end{tabular}
 \vspace{1ex} 
\end{table}

\label{lastpage}
\end{document}